\documentclass[aps,prb,twocolumn,superscriptaddress]{revtex4-1}

\pdfoutput=1

\usepackage[utf8]{inputenc}
\usepackage{graphicx}
\DeclareGraphicsExtensions{.png,.jpg,.pdf}

\usepackage{amsmath}
\usepackage{amssymb}

\usepackage{hyperref}
\hypersetup{colorlinks=true,linkcolor=blue,citecolor=blue}
\usepackage{float}

\usepackage[dvipsnames]{xcolor}

\usepackage{braket}
\usepackage{bm}
\usepackage{bbold}
\usepackage{siunitx}
\usepackage{chemformula}
\usepackage{comment}
\usepackage{cancel}
\usepackage{interval}

\begin{document}

\title{Optical orientation with linearly polarized light in transition metal dichalcogenides}

\author{G. Catarina}
\thanks{goncalo.catarina@inl.int}
\affiliation{QuantaLab, International Iberian Nanotechnology Laboratory, 4715-330 Braga, Portugal}

\author{J. Have}
\affiliation{Department of Materials and Production, Aalborg University, DK-9220 Aalborg East, Denmark}
\affiliation{Department of Mathematical Sciences, Aalborg University, DK-9220 Aalborg East, Denmark}

\author{J. Fern\'{a}ndez-Rossier}
\thanks{On leave from Departamento de F\'{i}sica Aplicada, Universidad de Alicante, 03690 San Vicente del Raspeig, Spain.}
\affiliation{QuantaLab, International Iberian Nanotechnology Laboratory, 4715-330 Braga, Portugal}

\author{N. M. R. Peres}
\affiliation{QuantaLab, International Iberian Nanotechnology Laboratory, 4715-330 Braga, Portugal}
\affiliation{Centro de F\'{i}sica das Universidades do Minho e Porto and Departamento de F\'{i}sica and QuantaLab, Universidade do Minho, Campus de Gualtar, 4710-057 Braga, Portugal}

\date{\today}

\begin{abstract}
We study the optical properties of semiconducting transition metal dichalcogenide monolayers under the influence of strong out-of-plane magnetic fields, using the effective massive Dirac model.
We pay attention to the role of spin-orbit coupling effects, doping level and electron-electron interactions, treated at the Hartree-Fock level.
We find that optically-induced valley and spin imbalance, commonly attained with circularly polarized light, can also be obtained with linearly polarized light in the doped regime.
Additionally, we explore an exchange-driven mechanism to enhance the spin-orbit splitting of the conduction band, in n-doped systems, controlling both the carrier density and the intensity of the applied magnetic field. 
\end{abstract}

\maketitle

\section{Introduction}
The discovery of two-dimensional (2D) systems whose quasiparticles are described in terms of a Dirac theory~\cite{Novoselov2005} has been one of the major breakthroughs over the last two decades in condensed matter physics and has fuelled research in the area of 2D materials~\cite{Novoselov2016,Roldan2017}. 
Graphene, that features gapless Dirac cones in the neighborhood of the Fermi energy~\cite{Neto2009}, is a paradigmatic example. 
Interestingly, there are also 2D semiconductors that require a description through a massive Dirac equation~\cite{Xiao2012,Goerbig2014}, instead of a Schr\"{o}dinger-like model.
Whereas both Dirac and Schr\"{o}dinger theories would yield similar energy bands, their wave functions and linear response are distinct. 
The massive Dirac Hamiltonian comprises a finite Berry curvature that entails an unconventional Hall response~\cite{Xiao2007}. 
The Landau level spectrum of massive Dirac electrons features valley-dependent zeroth Landau levels aligned with either the valence or the conduction bands~\cite{Koshino2010}.
These properties are absent for Schr\"{o}dinger quasiparticles.

The effective picture in terms of a gapped Dirac Hamiltonian provides an unifying description of materials that, from the chemical point of view, are quite different. 
For instance, whereas for graphene the Dirac states are made of $p_z$ orbitals~\cite{Neto2009}, for transition metal dichalcogenides they are made of $d_{x^2-y^2}$ and $d_{xy}$ orbitals in the valence band and $d_ {z^2}$ in the conduction band~\cite{Xiao2012,Kosmider2013}.

In this work, we study the optical response of massive Dirac systems under the influence of applied out-of-plane magnetic fields. 
We focus on the case of transition metal dichalcogenide (TMD) monolayers, \ch{MX2}, where \ch{M}=\ch{Mo},\ch{W} and \ch{X}=\ch{S},\ch{Se}, whose magneto-optical properties have attracted considerable interest both from the experimental~\cite{Aivazian2015,Srivastava2015,Schmidt2016,Wang2017} and theoretical~\cite{Rose2013,Chu2014} side.
These direct band gap semiconductors are object of intense scrutiny because of their strong light-matter coupling~\cite{Mak2010,Splendiani2010}, strong spin-orbit interactions~\cite{Xiao2012,Kosmider2013}, rich excitonic effects~\cite{Berkelbach2013,Ugeda2014,Chaves2017,Wang2018} and potential applications in the emergent field of valleytronics~\cite{Zhang2014,Mak2018}.   
Nevertheless, our results can be easily adapted to other systems described by a massive Dirac equation, such as gapped graphene~\cite{Jiang2010,Pedersen2011}, silicene and related materials~\cite{Tabert2013} or antiferromagnetic honeycomb semiconductors~\cite{Li2013}.

The effects of orbital coupling to an external out-of-plane magnetic field, as well as spin-orbit interactions, are explicitly taken into account.
Electron-electron interactions are considered at the Hartree-Fock level, but electron-hole attraction and corresponding excitonic effects are left for a companion publication~\cite{Have2018}. 

The rest of this paper is organized as follows. 
In Section~\ref{section:Hamiltonian}, we introduce the physical system and its model Hamiltonian, which forms the basis for the whole work.
Section~\ref{section:formalism} contains the formalism used to calculate the magneto-optical properties, in particular the derivation of the electric susceptibility response function.
The analysis of the results is presented in Section~\ref{section:longitudinal}, for the longitudinal susceptibility, \ref{section:Hall}, for the transverse susceptibility, and \ref{section:circular}, for the response to circularly polarized light.
Section~\ref{section:exchange} is devoted to the calculation of the exchange self-energy corrections.
Additional technical details are provided in the Appendixes.

\section{Model Hamiltonian} 
\label{section:Hamiltonian}
\begin{figure}
 \includegraphics[width=\columnwidth]{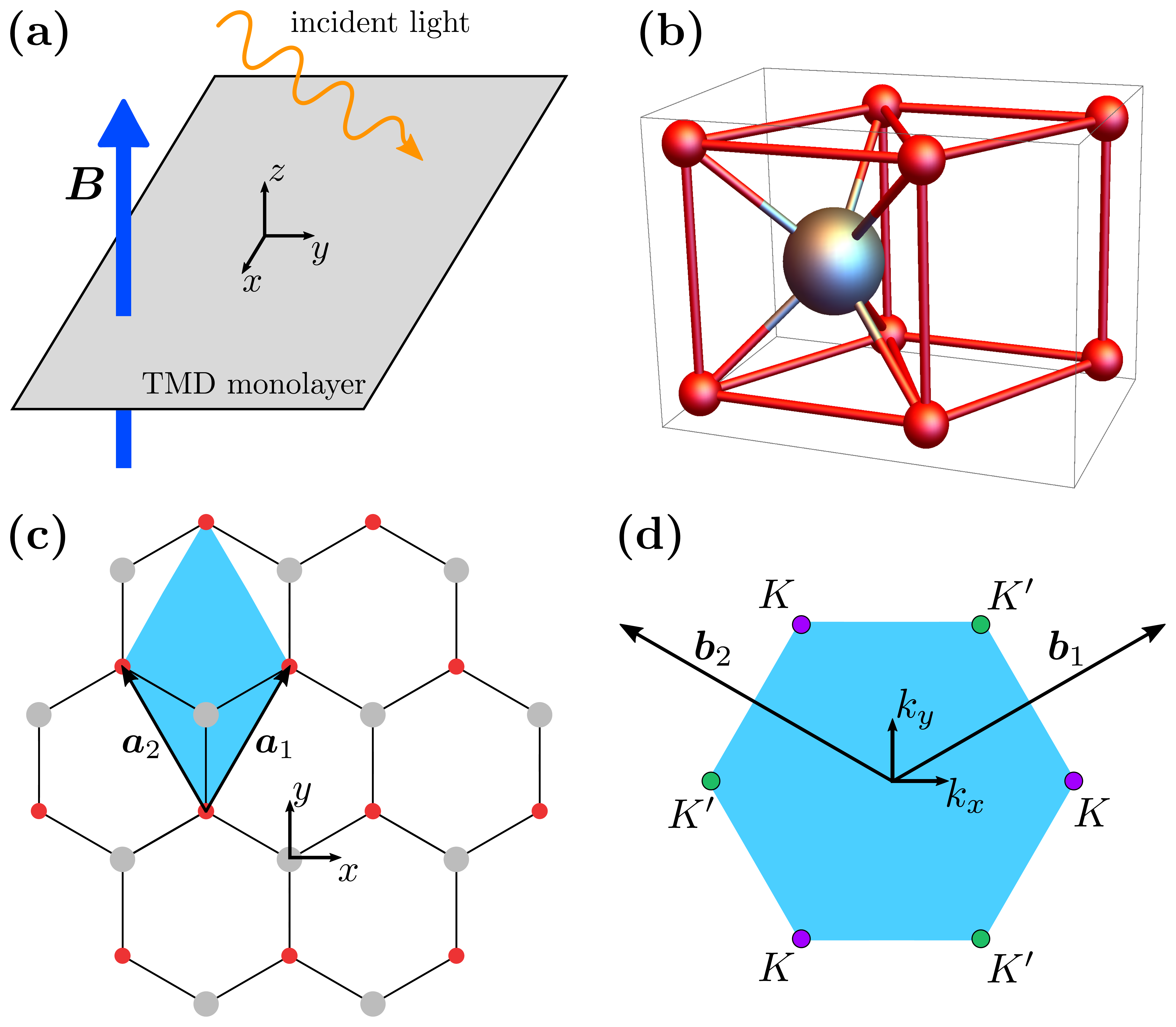}
 \caption{(Color online) Representation of the physical system. 
 (a): light is shinned into a transition metal dichalcogenide (TMD) monolayer subject to a perpendicular magnetic field, $\bm{B}$, uniform in space and time. 
 (b)-(c): the TMD crystal structure consists of an hexagonal lattice ---top view shown in  (c)--- of trigonal prismatic unit cells, (b), each of them containing one transition metal atom (big gray spheres) and two chalcogens (small red spheres); in (c), the blue region marks the unit cell of the crystal, defined by the primitive vectors $\bm{a}_1$ and $\bm{a}_2$.
 (d): corresponding (hexagonal) Brillouin zone, defined in reciprocal space by the primitive vectors $\bm{b}_1$ and $\bm{b}_2$, with the Dirac points $K$ and $K'$ indicated.}
 \label{fig:system}
\end{figure}

We consider a single-layer TMD in the $xy$-plane with a perpendicular uniform magnetic field pointing in the $z$-direction. 
The crystal structure consists of an hexagonal lattice of trigonal prismatic unit cells, each of them containing one transition metal atom and two chalcogens.
The resulting hexagonal Brillouin zone has two inequivalent sets of three equivalent corners, the so-called $K$ and $K'$ valleys (or Dirac points).
Due to the absence of an inversion center, the valley index provides an additional discrete degree of freedom for carriers in this system.
The physical system is depicted in Fig.~\ref{fig:system}.

In the low-energy regime, the electronic properties of TMD monolayers are often described by a massive Dirac Hamiltonian around the valleys~\cite{Xiao2012,Liu2013,Rose2013,Kormanyos2015}.
Spin-orbit coupling (SOC) splits both the valence and conduction bands, with opposite spin splittings at the two valleys, preserving time reversal symmetry thereby and leading to the so-called spin-valley coupling~\cite{Xiao2012}.
The magnitude of SOC splitting in the valence and conduction bands is different, on account of their different atomic orbital breakdown. 
The spin splitting of the valence band is of the order of hundreds of $\si{\milli\electronvolt}$ whereas, in the conduction band, it is smaller than few tens of $\si{\milli\electronvolt}$~\cite{Liu2013}.
Moreover, different TMD materials yield different relative signs of spin splitting in the conduction and valence bands at a given valley~\cite{Liu2013}.
In these systems, SOC commutes with the spin operator $S_z$. 
As a result, it can be introduced in a phenomenological manner~\cite{Ochoa2013,Chaves2017} by redefining the Dirac mass, including a valley ($\tau$) and spin ($s$) dependency, $\Delta \rightarrow \Delta_{\tau s}$, and adding an offset energy term, $\xi_{\tau s}$, defined below. 

In the presence of a uniform out-of-plane magnetic field, $\bm{B}=B_0 \hat{\bm{z}}$, the single-particle Hamiltonian for each valley and spin subspace is thus written, in the Landau gauge, as
\begin{equation}
H_0^{\tau,s} = v_F \left( \tau \sigma_x p_x + \sigma_y p_y + e B_0 x \sigma_y \right) + \Delta_{\tau s} \sigma_z + \xi_{\tau s} \mathbb{1}_2,
\label{eq:H0}
\end{equation}
where $\tau=\pm$ ($+$ for the $K$ valley and $-$ for the $K'$), $s=\uparrow(+), \downarrow(-)$, $v_F$ is the Fermi velocity, $\sigma_i (i=x,y,z)$ are the Pauli matrices with eigenvalues $\pm 1$, $\bm{p} = (p_x, p_y) = -\mathrm{i}\hbar \bm{\nabla}$ is the canonical electron momentum ($\hbar$ is the reduced Planck constant), $-e<0$ is the electron charge and $\mathbb{1}_2$ is the $2 \times 2$ identity matrix.  
The Pauli matrices and the identity matrix act on the space of the highest energy valence and lowest energy conduction states~\cite{Xiao2012}. 
The explicit forms of the valley- and spin-dependent Dirac mass, $\Delta_{\tau s}$, and offset energy, $\xi_{\tau s}$, read~\cite{Ochoa2013,Chaves2017}
\begin{equation}
\Delta_{\tau s} = \Delta - \tau s \frac{\Delta_{\text{SOC}}^\mathcal{V}-\Delta_{\text{SOC}}^\mathcal{C}}{4}, \quad 
\xi_{\tau s} = \tau s \frac{\Delta_{\text{SOC}}^\mathcal{V}+\Delta_{\text{SOC}}^\mathcal{C}}{4},
\end{equation}
where $\Delta_{\text{SOC}}^{\mathcal{V}}$ ($\Delta_{\text{SOC}}^{\mathcal{C}}$) is the spin splitting in the valence (conduction) band.
For $B_0=0$, the band gap is given by $2 \Delta_{\tau s}$.

The effective Hamiltonian, Eq.~\eqref{eq:H0}, shows that the dependency of the mass term on the valley and spin indexes is encoded in the product $\tau s$.   
In addition, the valley index appears on its own in the kinetic term, leading to valley-selective circular dichroism (introduced in Section~\ref{subsection:formalism_circular}), as we discuss in Section~\ref{section:circular}.
We neglect Zeeman splitting, that could be easily added as an additional term $g \mu_B B_0 \frac{s}{2} \mathbb{1}_2$, where $g$ is the $g$-factor and $\mu_B$ the Bohr magneton.
This term would split the energy bands of the two spin channels by $|g| \mu_B B_0 \simeq 0.12 B_0[\si{\tesla}] \si{\milli\electronvolt}$.
Compared to the spin splitting driven by the strong SOC, this effect is, for any reasonable scenario, negligible in the valence bands of TMDs.
As for the conduction bands, even though Zeeman and SOC can yield comparable magnitudes for strong applied fields, the results discussed in this paper are not substantially affected by the absence of Zeeman splitting in the model.
The effect of higher than first order $\bm{k} \cdot \bm{p}$ terms in the Hamiltonian~\cite{Kormanyos2015} has also been ignored.  

Closed analytical expressions for the eigenstates of $H^{\tau,s}_0$ can be obtained in terms of Landau levels that fall into two categories: the zeroth Landau level and the $n \neq 0$ Landau levels~\cite{Jiang2010,Koshino2010,Lado2013}.
The eigenvalues read 
\begin{equation}
E_{n,\lambda}^{\tau, s} = \lambda \sqrt{\Delta_{\tau s}^2 + \frac{1}{2} \left( \hbar \omega_0 \right)^2 n } + \xi_{\tau s}, 
\label{eq:spectrum}
\end{equation}
where $\frac{\omega_0}{2} = \frac{v_F}{l_B}$ is the characteristic angular frequency ($l_B = \sqrt{\frac{\hbar}{e B_0}}$ is the magnetic length) and $\{n;\lambda\}$ is the set of quantum numbers that describes the energy levels of this system, in which $n$ is the Landau level (LL) index and $\lambda$ the conduction ($\mathcal{C}$) or valence ($\mathcal{V}$) band index.
For the $n \neq 0$ LLs, $n=1,2,...$ and $\lambda=+(\mathcal{C}),-(\mathcal{V})$; the zeroth Landau level (0LL) is obtained setting $n=0$ and $\lambda=-\tau$.
The corresponding wave functions yield
\begin{equation}
\psi_{n,\lambda,k_y}^{\tau,s} (u,y) = \frac{\mathrm{e}^{\mathrm{i} k_y y}}{\sqrt{L_y}} \frac{\mathrm{e}^{-u^2/2}}{\sqrt{\sqrt{\pi} l_B}} C_{n,\lambda}^{\tau, s} 
\begin{pmatrix} 
\tilde{H}_{n_\tau} (u) \\ 
\mathrm{i} B_{n,\lambda}^{\tau, s} \tilde{H}_{n_\tau + \tau} (u) 
\end{pmatrix} ,
\end{equation}
where $k_y$ stands for the wave vector in the $y$-direction, which is quantized as $k_y = \frac{2\pi n_y}{L_y} , \ n_y \in \mathbb{Z}$ by applying periodic boundary conditions along the $y$-direction to a sample of length $L_y$.
We have also defined $u \equiv \frac{x}{l_B} + l_B k_y$, $n_\tau \equiv n-\frac{1+\tau}{2}$, $\tilde{H}_n \equiv \frac{1}{\sqrt{2^n n!}} H_n$ for $n\geq0$ (where $H_n$ are the Hermite polynomials) and $\tilde{H}_{-1} \equiv 0$.
The normalization constants, $C_{n,\lambda}^{\tau, s}$ and $B_{n,\lambda}^{\tau, s}$, are given by
\begin{equation}
C_{n,\lambda}^{\tau, s} = \sqrt{\frac{\bar{\Delta}_{\tau s} \left( \bar{\Delta}_{\tau s} + \check{E}^{\tau, s}_{n,\lambda} \right) + n}{\bar{\Delta}_{\tau s} \left( \bar{\Delta}_{\tau s} + \check{E}^{\tau, s}_{n,\lambda} \right) + 2n}} \in \mathbb{R}
\end{equation}
and
\begin{equation}
B_{\text{0LL}}^{\tau, s} = -\mathrm{i}, \quad
B_{n\neq 0,\lambda}^{\tau, s} = \frac{\sqrt{2n}}{ \bar{\Delta}_{\tau s} + \check{E}^{\tau, s}_{n,\lambda}} \in \mathbb{R},
\end{equation}
in which $\bar{\Delta}_{\tau s} \equiv \frac{2 \Delta_{\tau s}}{\hbar \omega_0}$, $\check{E}^{\tau, s}_{n,\lambda} \equiv \bar{E}^{\tau, s}_{n,\lambda} - \bar{\xi}_{\tau s}$, $\bar{E}^{\tau, s}_{n,\lambda} \equiv \frac{2 E^{\tau, s}_{n,\lambda}}{\hbar \omega_0}$ and $\bar{\xi}_{\tau s} \equiv \frac{2 \xi_{\tau s}}{\hbar \omega_0}$.

\begin{figure}[t]
 \includegraphics[width=\columnwidth]{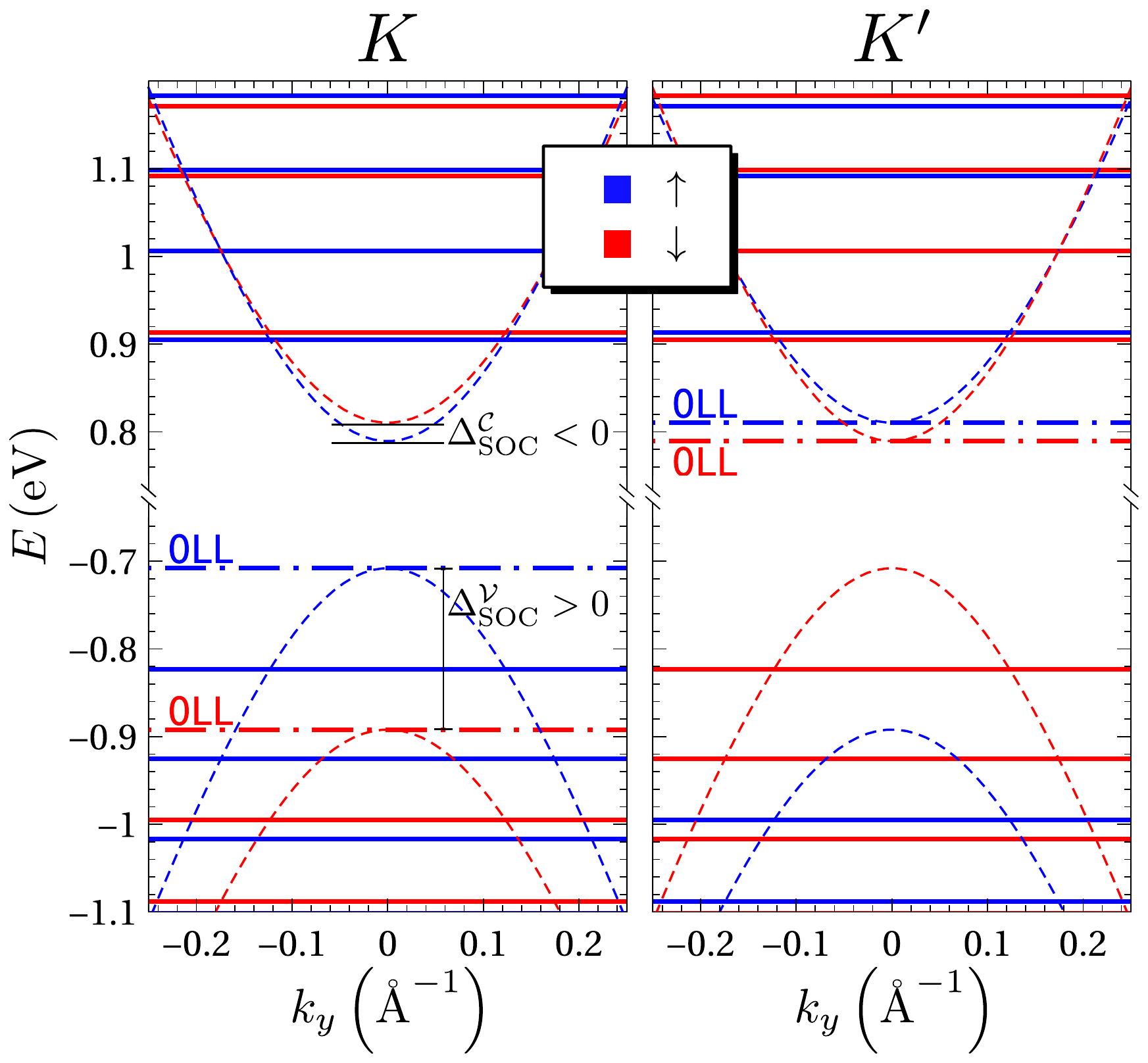}
 \caption{(Color online) Energy bands of monolayer \ch{MoSe2} in the Dirac approximation. 
 Different colors represent different spin projections: blue for spin up and red for spin down. 
 Dashed lines describe the solutions without external fields; band crossing exists in the conduction bands because $\Delta_{\text{SOC}}^\mathcal{C}<0$. 
 The application of an out-of-plane magnetic field ---$B_0 = 500 \si{\tesla}$ in this figure--- leads to the quantization of these bands into the Landau levels (horizontal lines); the unfeasible magnitude of $B_0$ is set only for readability purposes, as the observed features do not change qualitatively when working with practical values.
 Comparing the energy bands of both $K$ and $K'$ valleys, only the spin projection is interchanged, except for the zeroth Landau levels (dash-dotted lines).}
 \label{fig:energy_spectrum}
\end{figure}

The band structure implied by Eq.~\eqref{eq:spectrum} is depicted in Fig.~\ref{fig:energy_spectrum} for the case of \ch{MoSe2}.
Except for Section~\ref{section:exchange}, typical general values $\hbar v_F=3.5 \si{\electronvolt \angstrom}$ and $\Delta=0.8 \si{\electronvolt}$~\cite{Xiao2012} are fixed throughout the paper.
Regarding the SOC parameters, each TMD is treated in separate as there are significant differences among different materials, for instance on the sign of $\Delta_{\text{SOC}}^\mathcal{C}$.
The SOC values used in this work are listed in Table~\ref{tab:parameters}.

\begin{table}
  \begin{tabular}{l | c c c c }
      & $\Delta_{\text{SOC}}^\mathcal{V} \left( \si{\electronvolt} \right)$
      & $\Delta_{\text{SOC}}^\mathcal{C} \left( \si{\electronvolt} \right)$ 
      & $\tilde{\mu}_B^{(\tau s = +)} \left(\mu_B \right)$
      & $\tilde{\mu}_B^{(\tau s = -)} \left(\mu_B \right)$
      \\ \hline
      \ch{MoS2} & $0.148$ & $-0.003$ & $2.11$ & $1.92$ \\
      \ch{WS2} & $0.430$ & $+0.029$ & $2.30$ & $1.79$ \\
      \ch{MoSe2} & $0.184$ & $-0.021$ & $2.15$ & $1.89$ \\
      \ch{WSe2} & $0.466$ & $+0.036$ & $2.32$ & $1.77$ \\
  \end{tabular}
  \caption{List of spin-orbit coupling (SOC) parameters, $\Delta_{\text{SOC}}^{\mathcal{V}/\mathcal{C}}$, and effective Bohr magnetons, $\tilde{\mu}_B^{(\tau s)}$ (in units of Bohr magneton $\mu_B$), for different transition metal dichalcogenide materials.
  The SOC parameters are taken from Ref.~\onlinecite{Liu2013}.
  The effective Bohr magnetons are calculated through the expression defined in the text.
  By definition, $\tilde{\mu}_B^{(\tau s)}$ depends on the product of valley ($\tau$) and spin ($s$).}
  \label{tab:parameters}
\end{table}

The properties of the 0LL eigenstates are quite different from those of the $n \neq 0$ LLs.  
The energy levels of the $n \neq 0$ LLs depend on the product $\tau s$, meaning that we can correspond $K$ to $K'$ bands by interchanging the spin projections.
However, this does not hold for the $n=0$ LLs, whose energy is given by $E_{\text{0LL}}^{\tau, s} = -\tau \Delta_{\tau s} + \xi_{\tau s}$.
In fact, we see that the $K$ ($K'$) valley hosts a valence-like (conduction-like) 0LL spin doublet.
This doublet is split exclusively by SOC, as the 0LLs do not disperse with the applied magnetic field, which also contrasts with the $n \neq 0$ LLs.

It must be noted, however, that more elaborate calculations~\cite{Chu2014,Lado2016} reveal a valley-dependent spectrum that contrasts with the Dirac model.
Although the valley-dependent physics of the 0LL is captured in the same manner, these first-principles calculations show $n \neq 0$ LLs that are also different for both valleys, even when SOC is ignored~\cite{Lado2016}.

For most practical values of $n \neq 0$ and $B_0$, it is true that $\Delta_{\tau s}^2 \gg \frac{1}{2} \left( \hbar \omega_0 \right)^2 n$. 
Therefore, we can expand Eq.~\eqref{eq:spectrum} in Taylor series and obtain
\begin{equation}
E^{\tau s}_{n \neq 0,\mathcal{C}} \simeq \Delta + 2 \tilde{\mu}_B^{(\tau s)} n B _0 + \tau s \frac{\Delta_{\text{SOC}}^\mathcal{C}}{2}
\label{eq:Taylor_LLs_conduction}
\end{equation}
and
\begin{equation}
E^{\tau s}_{n \neq 0,\mathcal{V}} \simeq -\left(\Delta + 2 \tilde{\mu}_B^{(\tau s)} n B _0 - \tau s \frac{\Delta_{\text{SOC}}^\mathcal{V}}{2} \right),
\label{eq:Taylor_LLs_valence}
\end{equation}
where we have defined the effective Bohr magneton as $\tilde{\mu}_B^{(\tau s)} = \frac{e\hbar}{2 m_{\tau s}}$, in which $m_{\tau s} = \frac{\Delta_{\tau s}}{v^2_F}$ is the effective electron rest mass.
From these equations, it is clear that the $n \neq 0$ LLs disperse linearly with $n$ and $B_0$, but with a slope that is controlled by $\tilde{\mu}_B^{(\tau s)}$ and thus yields different values for $\tau s = +$ or $\tau s = -$ (see Table~\ref{tab:parameters}).
As a result, at a given valley, the sign of the spin splitting between two LLs with the same $n \neq 0$ and different spin $s$ can be reversed as we ramp either $n$ or $B_0$.  
This is apparent in the conduction bands of Fig.~\ref{fig:energy_spectrum} and is a direct consequence of the fact that SOC leads to a spin-dependent non-relativistic mass in the Dirac theory, which in turn controls LL dispersion.

\section{Magneto-optical response: formalism} 
\label{section:formalism}
In this section, we introduce a general formalism to calculate the magneto-optical response in metals and semiconductors: the equation of motion (EOM) method~\cite{Ferreira2011}, a technique based on Ref.~\onlinecite{Peres2010} and generalized to include the effect of external magnetic fields.
The EOM method permits to derive analytical expressions of response functions that are fully equivalent to the Kubo formula when linear response theory is employed and electron-electron interactions are not taken into account.
Here, we apply this formalism to the Hamiltonian described in Section~\ref{section:Hamiltonian} and derive, within the linear response regime, analytical expressions for the electric susceptibility tensor in the cartesian basis, which are then manipulated to explicitly address the case in which the incident light is circularly polarized.
Free carrier transitions are considered in a first approximation, disregarding all the Coulomb interactions and thus treating electrons and holes as quasi-free particles.
Compared to the Kubo formula, the advantage of the EOM method is that, by treating Coulomb effects at the same level of the interaction with light, further corrections can be introduced within the same formalism.
In Section~\ref{section:exchange}, we account for Coulomb interactions at the self-energy level.
The role of excitonic effects is the main subject of a forthcoming publication~\cite{Have2018}.

\subsection{Dipole matrix elements}
The interaction with light is included, within the dipole approximation, via the following Hamiltonian:
\begin{equation}
H_I = - \bm{d} \cdot \bm{\mathcal{E}} = e \bm{r} \cdot \bm{\mathcal{E}}(t),
\end{equation}
where $\bm{r} = (x,y)$ is the 2D position vector, $\bm{d} = -e \bm{r}$ is the electric dipole moment and $\bm{\mathcal{E}} = \bm{\mathcal{E}}(t)$ is the electric field of the incident light, which is assumed homogeneous and dependent of the time $t$.

The method used in this paper relies on the calculation of the expectation value of the electric polarization density operator with regard to the unperturbed Hamiltonian, whose (complete) basis is $\alpha=\{n;\lambda;k_y\}$. 
Therefore, the matrix elements of the polarization density created by the dipole, $\bm{P} = \frac{\bm{d}}{A}$ ($A$ is the area of the system), are relevant quantities that define optical selection rules.

The computation of the dipole matrix elements in each one of the $\eta=\{\tau;s\}$ subspaces, $\bm{d}^{\eta}_{\alpha \rightarrow \alpha'} = \braket{\alpha'|\bm{d}|\alpha}_\eta = \left( \bm{d}^{\eta}_{\alpha' \rightarrow \alpha} \right)^*$, shows that only transitions between the same $k_y$ are coupled, i.e., $\bm{d}^{\eta}_{\alpha \rightarrow \alpha'} = \delta_{k_y,k'_y} {\bm{d}^{\eta}}_{\{n;\lambda\}}^{\{n';\lambda'\}}$~\footnote{This result is easily obtained using that $\braket{\alpha'|\bm{r}|\alpha}_\eta = \frac{\braket{\alpha'|\left[ \bm{r},H^\eta_0 \right]|\alpha}_\eta}{E^\eta_\alpha-E^\eta_{\alpha'}} = \mathrm{i} \hbar v_F \frac{\braket{\alpha'|\left(\tau \sigma_x,\sigma_y \right)|\alpha}_\eta}{E^\eta_\alpha-E^\eta_{\alpha'}}$, for $\alpha \neq \alpha'$, followed by the spatial integration.}.
In addition, it also reveals that the only nonzero terms are
\begin{equation}
{\bm{d}^{\eta}}_{\{n;\lambda\}}^{\{n+\tau;\lambda'\}} = \frac{-e \hbar v_F}{E_{n,\lambda}^{\eta} - E_{n+\tau,\lambda'}^{\eta}} 
C_{n+\tau,\lambda'}^{\eta} C_{n,\lambda}^{\eta} B_{n,\lambda}^{\eta} \left( -\tau, \mathrm{i} \right),
\end{equation}
for $n+\tau \geq 0$, and
\begin{equation}
\text{\small $
{\bm{d}^{\eta}}_{\{n;\lambda\}}^{\{n-\tau;\lambda'\}} = \frac{-e \hbar v_F}{E_{n,\lambda}^{\eta} - E_{n-\tau,\lambda'}^{\eta}} 
C_{n-\tau,\lambda'}^{\eta} C_{n,\lambda}^{\eta} \left( B_{n-\tau,\lambda'}^{\eta} \right)^* \left( \tau, \mathrm{i} \right), 
$}
\end{equation}
for $n-\tau \geq 0$.
The former relations embody the following optical selection rule: for an electron with wave vector $k_y$ and in a given LL with index $n$, the absorption of a photon can only induce a transition ---which can be intra or interband--- to a state with the same wave vector and with a LL index given by $n' = n \pm 1 \geq 0$.
This well-known selection rule~\cite{Gusynin2007a,Pedersen2011,Ferreira2011,Tabert2013} adds up to the ones imposed by construction: the decoupling of the valleys, which is consistent with the dipole approximation, and the decoupling of the spins, which is consistent with the lack of spin-flip terms in the Hamiltonian.

\subsection{Electric susceptibility}
Moving to the Heisenberg picture, and introducing the (time-dependent) creation/annihilation fermionic operators in this representation, $\hat{c}^\dagger_{\alpha,\eta}(t)/\hat{c}_{\alpha,\eta}(t)$, the total Hamiltonian can be written as
\begin{equation}
\hat{H}(t) = \hat{H}_0(t) + \hat{H}_{I}(t),
\label{eq:Htotal}
\end{equation}
where
\begin{equation}
\hat{H}_0(t) = \sum_{\eta,\alpha} E^{\eta}_{\alpha} \hat{c}^\dagger_{\alpha,\eta}(t)\hat{c}_{\alpha,\eta}(t)
\end{equation}
is the unperturbed Hamiltonian and
\begin{equation}
\hat{H}_I(t) = -\bm{\mathcal{E}}(t) \cdot \sum_{\eta,\alpha,\alpha'} \bm{d}^{\eta}_{\alpha \rightarrow \alpha'} \hat{c}^\dagger_{\alpha',\eta}(t) \hat{c}_{\alpha,\eta}(t) 
\label{eq:HI}
\end{equation}
is the Hamiltonian that describes the dipole interaction with light.
Repeating the same procedure for the polarization density, we get
\begin{equation}
\hat{\bm{P}}(t) = \frac{1}{A} \sum_{\eta,\alpha,\alpha'} \bm{d}^{\eta}_{\alpha \rightarrow \alpha'} \hat{c}^\dagger_{\alpha',\eta}(t) \hat{c}_{\alpha,\eta}(t)
\end{equation}
and, defining the general operator $\hat{T}_{\alpha,\alpha'}^\eta (t) \equiv \hat{c}^\dagger_{\alpha',\eta}(t) \hat{c}_{\alpha,\eta}(t)$, whose EOM reads
\begin{equation}
-\mathrm{i} \hbar \frac{d}{dt} \hat{T}_{\alpha,\alpha'}^\eta (t) =  \left[ \hat{H}(t), \hat{T}_{\alpha,\alpha'}^\eta (t)\right], 
\label{eq:EOM}
\end{equation}
it is apparent that the time evolution of the polarization density operator can be achieved by solving Eq.~\eqref{eq:EOM}.

The details regarding the technical step of solving the above-mentioned EOM are provided in Appendix~\hyperref[sec:AppendixA]{A}. 
In short, we start by calculating the commutator, so we can explicitly write down the differential equation. 
Then, we solve for its expectation value within the linear response approximation and in the adiabatic regime. 
The outcome is the expression for $\braket{\hat{\bm{P}}(t)} \equiv \bm{P}(t)$ within the former approximations. 

Expressing $\bm{P}(t)$ through its Fourier transform, $\bm{P}(\omega)$, we are then able to recognize the (homogeneous and dynamical) electric susceptibility tensor, 
\begin{equation}
\chi (\omega) = 
\begin{pmatrix} 
\chi_{xx} (\omega) & \chi_{xy} (\omega) \\ 
\chi_{yx} (\omega) & \chi_{yy} (\omega)
\end{pmatrix}, 
\end{equation}
via the constitutive relation $\bm{P}(\omega) = \varepsilon_0 \chi(\omega) \bm{\mathcal{E}}(\omega)$, where $\varepsilon_0$ is the vacuum permittivity, $\omega$ is the angular frequency and $\bm{\mathcal{E}}(\omega)$ is the Fourier transform of $\bm{\mathcal{E}}(t)$.
Putting it all together, we conclude that $\chi_{xx} = \chi_{yy}$ and $\chi_{xy} = -\chi_{yx}$, which is an expected result for systems with $C_6$ symmetry~\cite{Nowick1995}.
The final expressions for the longitudinal and transverse susceptibility, $\chi_{xx}$ and $\chi_{yx}$ (respectively), read
\begin{equation}
\chi_{xx} (\omega) = \mathcal{S}_{+}(\omega) , \quad \chi_{yx} (\omega) = \mathrm{i} \mathcal{S}_{-}(\omega),
\label{eq:chi_1}
\end{equation}
where $\mathcal{S}_\pm(\omega)$ are auxiliar functions defined as
\begin{equation}
\begin{split}
& \text{\footnotesize $
\mathcal{S}_{\pm} (\omega) \equiv \sum_\eta \sum_{\{n;\lambda\},\lambda'} \frac{f\left(E_{n+1,\lambda'}^\eta\right) - f\left(E_{n,\lambda}^\eta\right)}{2 \pi l_B^2 \varepsilon_0}
\left| {d_x^{\eta}}_{\{n;\lambda\}}^{\{n+1;\lambda'\}} \right|^2 \times
$} \\
& \text{\footnotesize $
\quad \times \left( \frac{1}{E_{n,\lambda}^\eta - E_{n+1,\lambda'}^\eta + \hbar \omega + \mathrm{i} \Gamma} \pm 
\frac{1}{E_{n,\lambda}^\eta - E_{n+1,\lambda'}^\eta - \hbar \omega - \mathrm{i} \Gamma} \right),
$} 
\label{eq:chi_2}
\end{split}
\end{equation}
in which $\Gamma$ is a phenomenological parameter that accounts for disorder within the adiabatic approximation and $f$ stands for the Fermi-Dirac distribution at Fermi level $\mu$ and absolute temperature $T$ (see Appendix~\hyperref[sec:AppendixA]{A} for details).
Throughout this work, we have set $\Gamma=7\si{\milli\electronvolt}$, which is a rather low but feasible value that corresponds to samples of TMDs encapsulated in hexagonal boron nitride and with little impurity~\cite{Cadiz2017,Ajayi2017}.
The disorder parameter does not influence the results presented in this paper if the full width at half maximum of the lorentzian implicit in Eq.~\eqref{eq:chi_2}, $2\Gamma$, is smaller (or at least of the same order of magnitude) than the LL splitting, which is roughly given by $2 \tilde{\mu}_B^{(\tau s)} B_0 \sim 0.2 B_0[\si{\tesla}] \si{\milli\electronvolt}$.
This explains why we have set such strong (but still feasible) out-of-plane magnetic fields in the optical response results.
For clarity purposes, we stress that, to write $\chi_{yx} (\omega)$ in its final form, we have used that $\left( {d_y^{\eta}}_{\{n;\lambda\}}^{\{n+1;\lambda'\}}  \right)^* {d_x^{\eta}}_{\{n;\lambda\}}^{\{n+1;\lambda'\}} = \mathrm{i} \left| {d_x^{\eta}}_{\{n;\lambda\}}^{\{n+1;\lambda'\}} \right|^2$.

\subsection{Circularly polarized light} 
\label{subsection:formalism_circular}
Associated with the will of exploring valley-based optoelectronic applications, many studies deal with circularly polarized light~\cite{Yao2008,Cao2012}.
The underlying mechanism is valley-selective circular dichroism, i.e., differential absorption of left- and right-handed photons when comparing the contributions from inequivalent valleys.
This contrasts with the usual circular dichroism, for which there is a difference in the (overall) absorption of left-handed ($\sigma^-$) and right-handed ($\sigma^+$) light.
At $\bm{B}=\bm{0}$, the massive Dirac Hamiltonian breaks time reversal symmetry in each valley, leading to a circular dichroism that is valley-dependent~\cite{Yao2008,Ezawa2013,Xu2014}.
In this case, the total circular dichroism vanishes when summing over valleys, as time reversal symmetry is restored.
However, illumination with circularly polarized light results in populations of excited carriers with valley polarization.
Conceptually, this permits to access the valley pseudospin degree of freedom, the key idea of valleytronics.
In addition, because of the strong SOC, the same mechanism also leads to an optically-induced spin imbalance in TMD materials~\cite{Xu2014}.
In this work, we propose a complementary route to induce both valley and spin polarization in TMDs with linearly polarized light.
Nevertheless, for completeness, we discuss here the case of incident circularly polarized light, which is relevant for Section~\ref{section:circular}.

Assuming incident light with circular polarization, i.e., $\bm{\mathcal{E}}(\omega) = \bm{\mathcal{E}}^\pm(\omega) \equiv \frac{\mathcal{E}_0 (\omega)}{\sqrt{2}} \left(1, \mathrm{e}^{\pm \mathrm{i} \pi/2}\right)$, where $\mathcal{E}_0 (\omega)$ is the (equal) amplitude of the two plane waves and $\pm$ stands, in the point of view of the source, for right and left polarization, respectively, the electric susceptibility tensor is shown to be diagonal in the circular basis, with the diagonal elements given by
\begin{equation}
 \chi_\pm (\omega) = \chi_{xx} (\omega) \pm \mathrm{i} \chi_{yx} (\omega). 
 \label{eq:chi_cartesian_circular}
\end{equation}
This relation lays on symmetry foundations as it is valid as long as $\chi_{xx} = \chi_{yy}$ and $\chi_{xy} = -\chi_{yx}$ are satisfied.
Moreover, it shows that circular dichroism is encoded in the real part of $\chi_{yx}$.

\section{Longitudinal susceptibility} 
\label{section:longitudinal}
We now move onto the discussion of the main features that characterize the low-energy non-interacting magneto-optical response in TMDs.
Although Coulomb interactions are known to be significant~\cite{Aivazian2015,Wang2017,Wang2018}, the study of the non-interacting limit provides reference for further analyses.

In this section, we discuss the results for the dynamical longitudinal susceptibility, $\chi_{xx} (\omega)$. 
This quantity is directly relevant in modeling experiments where TMDs are excited with linearly polarized light.
In addition, $\chi_{xx} (\omega)$ contributes to $\chi_\pm (\omega)$, as seen in Eq.~\eqref{eq:chi_cartesian_circular}.
Therefore, it is also important to interpret the response to circularly polarized light (Section~\ref{section:circular}).

The evaluation of Eq.~\eqref{eq:chi_2} requires a cutoff, as usual when dealing with low-energy effective models.  
For this matter, we establish a range of frequencies that is consistent with the underlying $\bm{k} \cdot \bm{p}$ theory that leads to the Dirac Hamiltonian.
By construction, this theory is only valid in the neighborhood of the high-symmetry $K$ and $K'$ points, which sets an energy window out of which the model does not work.  
Taking a energy window of $\interval{-1.5}{1.5}~\si{\electronvolt}$ ---for which the upper bound lies $\sim 0.7\si{\electronvolt}$ above the bottom of the conduction band--- and bearing in mind the optical selection rules, plus the Pauli exclusion principle, we see that $\hbar \omega \lesssim 3\si{\electronvolt}$ is a suitable criterion, as it contemplates all and only the transitions between bands within the energy window.
This provides an intrinsic cutoff for the imaginary part of $\chi_{xx} (\omega)$, given that the only bands that contribute satisfy $\left| E_{n,\lambda}^\eta - E_{n+1,\lambda'}^\eta \right| \simeq \hbar \omega$.
For the real part, we have found that numerical convergence is attained with a cutoff energy of $|E_\text{cut}| \sim 4 \si{\electronvolt}$, which corresponds to a cutoff in the LLs, $n_\text{cut}$, that varies roughly as $4\times10^4 \left(B_0[\si{\tesla}]\right)^{-1}$.

The analysis of the results in this section is divided into three main categories that depend on the doping level.
We first consider the case of an intrinsic TMD, with $\mu$ lying inside the gap.
Then, we focus on the doped regime and separate two distinct scenarios.
First, we take a system on which the 0LLs do not participate in the optical transitions.
Second, we discuss the case of a TMD n-doped (p-doped) up to the first 0LL in the conduction (valence) band, for which the optical transitions that involve the 0LLs take a predominant role.

\subsection{Undoped regime: Fermi level in the gap} 
\label{subsection:longitudinal_undoped}
\begin{figure}
 \includegraphics[width=\columnwidth]{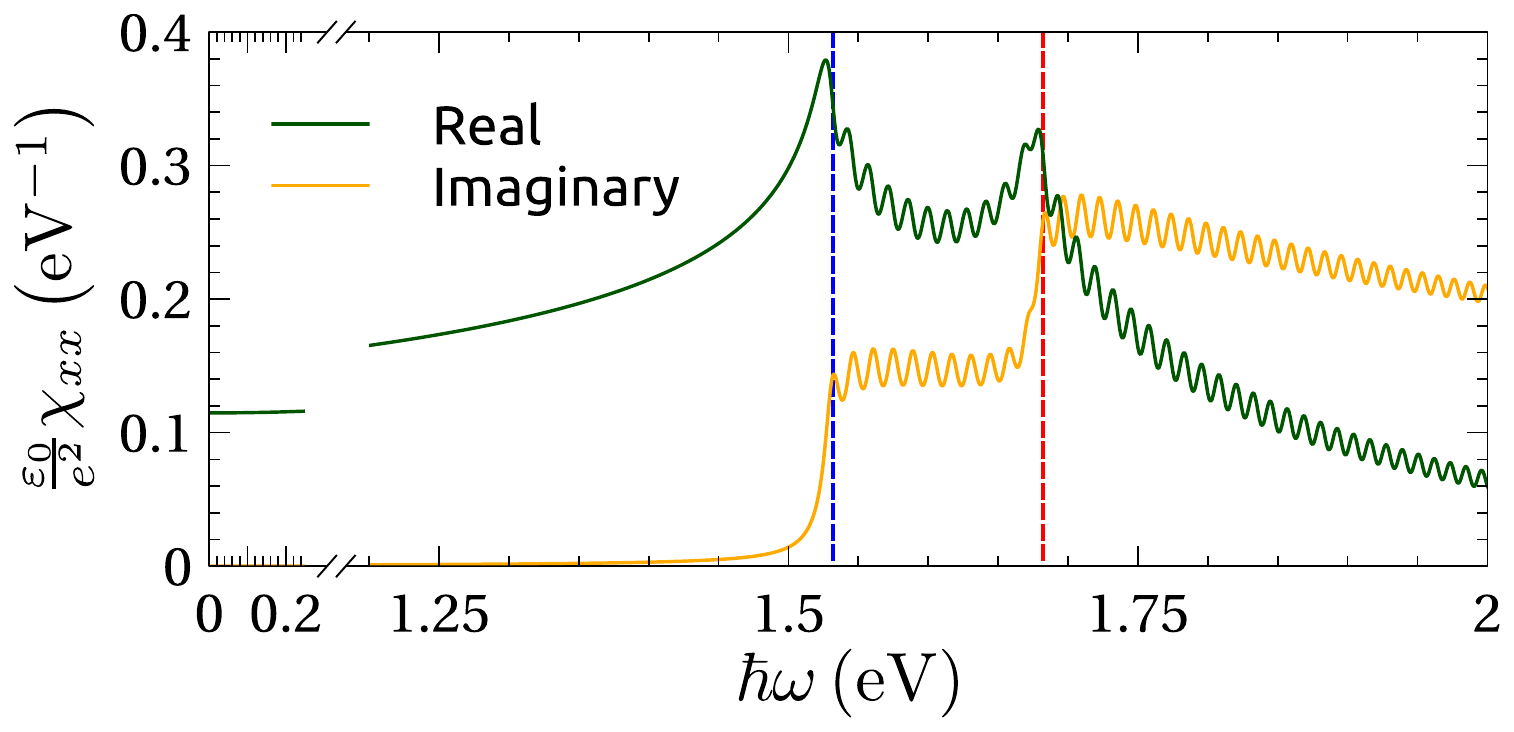}
 \caption{(Color online) Longitudinal susceptibility, $\chi_{xx}$, as a function of the photon energy, in monolayer \ch{MoS2} at the charge neutrality point and for a magnetic field of $30\si{\tesla}$ (results independent of the temperature). 
 The imaginary part, which is directly related with optical absorption, shows a sequence of peaks that correspond to the allowed optical transitions. 
 The vertical dashed lines mark the energy of the less energetic transition for each spin-valley product: for the $K$ ($K'$) valley, blue is for spin up (down) and red for spin down (up).
 The presence of a plateau between the vertical lines is the signature of spin-orbit coupling effects.}
 \label{fig:chixx_mugap}
\end{figure}

As we discuss in Section~\ref{section:Hall}, $\chi_{yx}$ vanishes for arbitrary $\omega$ in the undoped regime. 
Thus, for intrinsic TMDs, the magneto-optical response is governed exclusively by $\chi_{xx}$.  
When $\mu$ lies in the gap, intraband transitions are Pauli blocked, as thermal activations are negligible compared to the band gap, even at room temperature ($k_B T \simeq 26\si{\milli\electronvolt}$ for $T=300\si{\kelvin}$, compared to gaps in the order of $2\Delta = 1.6\si{\electronvolt}$).  
Therefore, in the undoped regime, the magneto-optical response is independent of the temperature and fully driven by interband transitions.
Fig.~\ref{fig:chixx_mugap} shows a plot of $\chi_{xx} (\omega)$ in a neutral \ch{MoS2} for $B_0 = 30 \si{\tesla}$, whose discussion follows below.

The imaginary part of $\chi_{xx} (\omega)$ describes photon absorption processes, induced when the photon energy matches the energy difference between an occupied and an empty state.  
The resulting curve features a structure of peaks that correspond to interband transitions satisfying the optical selection rules, which are summarized in Table~\ref{tab:transitions}.
It must be noted that, although spin-valley coupling is not manifest in the LL spectrum due to the valley-dependent 0LLs (see Fig.~\ref{fig:energy_spectrum}), $\tau s$ is still a relevant quantity to characterize transition energies, as all of them are maintained when we change valley and spin at the same time, even if the 0LLs are involved. 

\begin{table}
  \begin{tabular}{l | c | c c}
    & $K,s$ & $K',-s$  \\ \hline
    $\mathcal{T}^{(\tau s)}_0$ & $ \{0;\mathcal{V}\} \rightarrow \{1;\mathcal{C}\}$ & $\{1;\mathcal{V}\} \rightarrow \{0;\mathcal{C}\}$\\
    \hline
    $\mathcal{T}^{(\tau s)}_1$ & $\{1;\mathcal{V}\} \rightarrow \{2;\mathcal{C}\}$ & $\{2;\mathcal{V}\} \rightarrow \{1;\mathcal{C}\}$\\
    & $\{2;\mathcal{V}\} \rightarrow \{1;\mathcal{C}\}$ & $\{1;\mathcal{V}\} \rightarrow \{2;\mathcal{C}\}$\\
    \hline
    $\mathcal{T}^{(\tau s)}_2$ & $\{2;\mathcal{V}\} \rightarrow \{3;\mathcal{C}\}$ & $\{3;\mathcal{V}\} \rightarrow \{2;\mathcal{C}\}$\\
    & $\{3;\mathcal{V}\} \rightarrow \{2;\mathcal{C}\}$ & $\{2;\mathcal{V}\} \rightarrow \{3;\mathcal{C}\}$\\
    \hline
    \hline
   $\mathcal{T}^{(\tau s)}_{n>0}$ & $\{n;\mathcal{V}\} \rightarrow \{n+1;\mathcal{C}\}$ & $\{n+1;\mathcal{V}\} \rightarrow \{n;\mathcal{C}\}$\\
    & $\{n+1;\mathcal{V}\} \rightarrow \{n;\mathcal{C}\}$ & $\{n;\mathcal{V}\} \rightarrow \{n+1;\mathcal{C}\}$
  \end{tabular}
  \caption{List of the allowed optical transitions in intrinsic transition metal dichalcogenides, organized by their energies ($\mathcal{T}^{(\tau s)}_0, \mathcal{T}^{(\tau s)}_1, ...$).
  The representation of the transitions that correspond to each energy is separated by valley $\tau$, for a fixed spin-valley product (in this case given by $\tau s = s$, where $s$ is the spin index).
  There are four degenerate transitions for every energy, except for $\mathcal{T}^{(\tau s)}_0$, for which there are two.
  Transitions with equal contributions to the optical response are presented in the same line.}
  \label{tab:transitions}
\end{table}

Within the frequency range $\mathcal{T}^{(\tau s = +)}_0 < \hbar \omega < \mathcal{T}^{(\tau s = -)}_0$, where $\mathcal{T}^{(\tau s = \pm)}_0 = E^{K,\pm}_{1,\mathcal{C}} - E^{K,\pm}_\text{0LL} = E^{K',\mp}_\text{0LL} - E^{K',\mp}_{1,\mathcal{V}}$ are the transition energies that correspond to the vertical blue and red lines in Figure~\ref{fig:chixx_mugap} (respectively), only two (out of four) flavors of $\tau$ and $s$ contribute to the absorption, namely the ones that respect $\tau s = +$.
For $\hbar \omega > \mathcal{T}^{(\tau s = -)}_0$, the absorption curve features a second step that marks the entrance of transitions with $\tau s = -$.
The energy splitting of the two thresholds, given by $\mathcal{T}^{(\tau s = -)}_0 - \mathcal{T}^{(\tau s = +)}_0$, depends explicitly on the SOC parameters and is easily shown to vanish if and only if $\Delta_{\text{SOC}}^\mathcal{V} = \Delta_{\text{SOC}}^\mathcal{C} = 0$.
Thus, the presence of a plateau in $\text{Im}\{ \chi_{xx} (\omega) \}$ is a direct consequence of SOC interactions. 

We now discuss the intensity of the degenerate transitions, which come in doublets for $\mathcal{T}^{(\tau s)}_0$ and in quadruplets for all the other transition energies, as depicted in Table~\ref{tab:transitions}.
The height of the transitions is governed by the dipole matrix elements in Eq.~\eqref{eq:chi_2}, which satisfy the identity
\begin{equation}
\left| {d_x^{\tau,s}}_{\{n;\lambda\}}^{\{n+1;\lambda'\}} \right|^2 = \left| {d_x^{-\tau,-s}}_{\{n;-\lambda\}}^{\{n+1;-\lambda'\}} \right|^2.
\label{eq:dipoles_ident}
\end{equation}
This relation shows that ``counterpart transitions'', i.e., transitions with the same energy and equal contributions to the optical response, are obtained by changing valley, spin and also the band indexes at the same time.
In Table~\ref{tab:transitions}, we present the counterpart transitions in the same line.
It is therefore clear that every absorption peak in Fig.~\ref{fig:chixx_mugap} (which is characterized by a given $\tau s$ product), has equal contributions from the two possible $\tau$ and $s$ combinations.
For instance, using the notation of Table~\ref{tab:transitions}, this means that a peak with energy $\mathcal{T}^{(\tau s = +)}_n$ has equal contributions from $\tau=K, s=\uparrow$ and $\tau=K', s=\downarrow$.

Interestingly, in the case of the quadruplets, the two pairs of counterpart transitions are not equivalent in intensities.
In fact, the computation of the dipole matrix elements shows that one pair of transitions is overwhelmingly stronger than the other.
This feature cannot be observed through the spin and valley breakdown of the absorption curve because both the weak and strong pairs of transitions are allowed in the undoped regime.
However, as we discuss in Section~\ref{subsection:longitudinal_doped1}, doping allows to explore this property.

The real part of $\chi_{xx} (\omega)$, which describes the reactive dielectric response of the TMD, is also shown in Fig.~\ref{fig:chixx_mugap}. 
Expectedly, for in-gap frequencies, it decays smoothly as we decrease $\hbar \omega$ below the absorption threshold.
Above the absorption threshold, it oscillates as a function of the frequency, due to the presence of many resonant peaks in absorption.

\subsection{Doped system with optical transitions to zeroth Landau levels Pauli blocked} 
\label{subsection:longitudinal_doped1}
Away from charge neutrality, we find two fundamental differences with the undoped regime.
First, intraband transitions enter into play, while some of the interband ones become Pauli blocked.
Second, the AC Hall response, given by $\chi_{yx} (\omega)$, is no longer null, as we explore in Section~\ref{section:Hall}. 
The carrier density implied to get to this regime can arise either from gating or chemical doping.

\begin{figure}
 \includegraphics[width=\columnwidth]{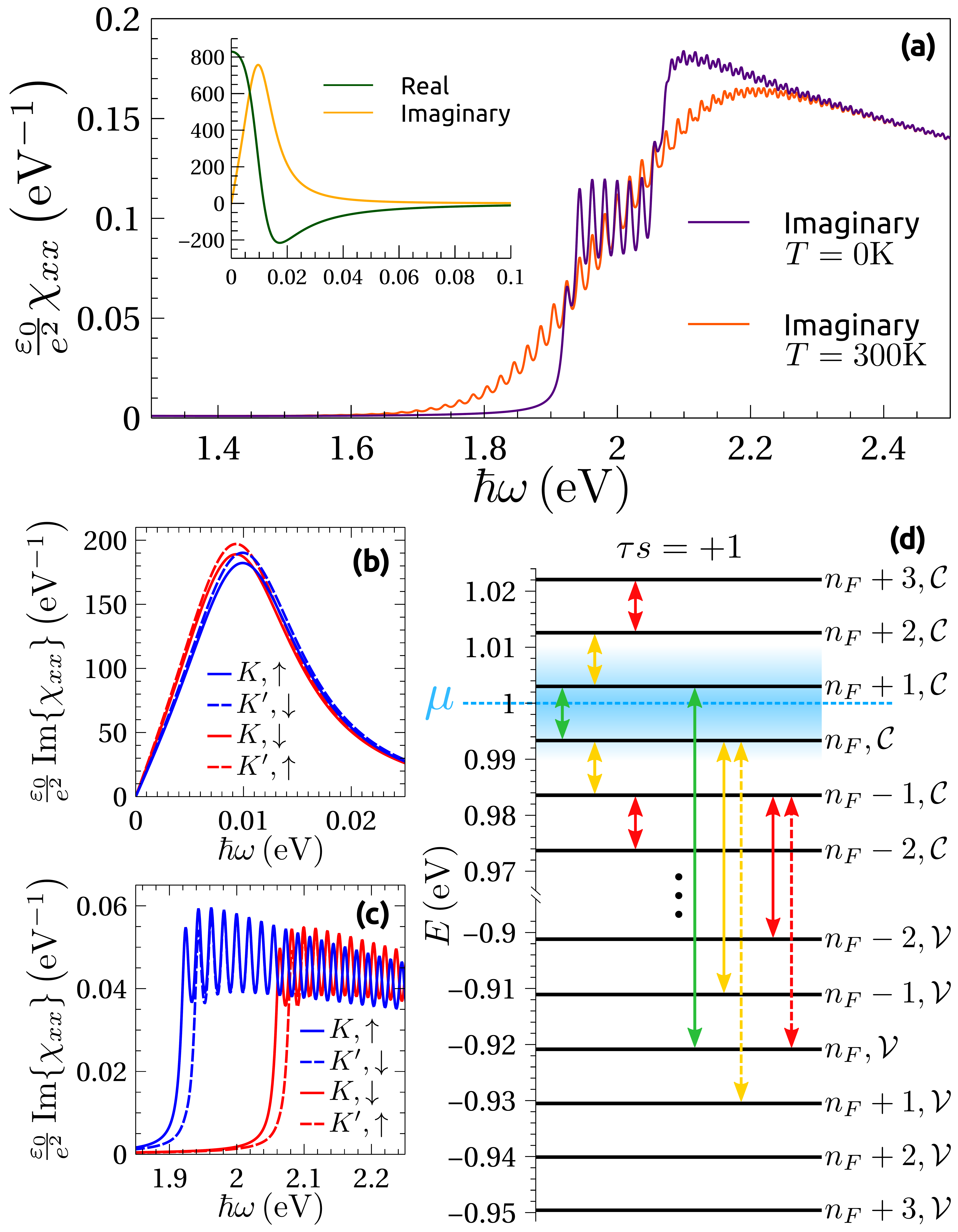}
 \caption{(Color online) Longitudinal magneto-optical response in a doped (Fermi level $\mu=1\si{\electronvolt}$) monolayer \ch{MoS2}, for a magnetic field of $50\si{\tesla}$: in (a), the longitudinal susceptibility, $\chi_{xx}$, is plotted as a function of the photon energy (results in the inset are roughly independent of the temperature $T$); (b) and (c) show the valley and spin breakdown of the absorptive part of $\chi_{xx}$ at zero absolute temperature; in (d), a scheme of the optical transitions between the energy bands is presented. 
 Discussion is provided in the text.}
 \label{fig:chixx_doped_+schemes}
\end{figure}

We start with the case where the 0LLs cannot participate in the optical transitions, neither as initial nor final states. 
Due to the optical selection rules, it suffices to have $\mu$ lying above (below) both $n=1$ LLs in the conduction (valence) band.
In this regime, the system is a quantum Hall insulator and the ground state has no spin nor valley polarization.
Without loss of generality, we take the example of a n-doped \ch{MoS2}, with $\mu = 1\si{\electronvolt}$ ($\sim 0.2\si{\electronvolt}$ above the bottom of the conduction band), for a magnetic field of $50\si{\tesla}$.
The overview of the results is presented in Fig.~\ref{fig:chixx_doped_+schemes}, and its analysis follows below.

Intra and interband absorption occur at very different frequencies, as observed in Fig.~\ref{fig:chixx_doped_+schemes}-(a).
The energy scale of the intraband absorption peak is controlled by the energy difference between two adjacent LLs in the same band, which, using Eqs.~\eqref{eq:Taylor_LLs_conduction} and \eqref{eq:Taylor_LLs_valence}, can be estimated as $2 \tilde{\mu}_B^{(\tau s)} B _0 \sim 0.2 B_0 [\si{\tesla}] \si{\milli\electronvolt}$.
Even for a very large field of $50\si{\tesla}$, we see that the intraband peak occurs around $\hbar \omega = 10\si{\milli\electronvolt} \ll 2\Delta$.
Thus, the discussion of the intra and interband parts of the magneto-optical spectrum can be separated.

At $T=0$, the intraband peak in absorption has contributions from a total of four transitions.
These intraband transitions connect the last occupied LL, $\{n;\lambda\} = \{n_F;\text{sign}(\mu)\}$, and the first empty one, $\{n;\lambda\} = \{n_F+\text{sign}(\mu);\text{sign}(\mu)\}$, for the four channels of $\tau$ and $s$.
Due to spin-valley coupling, the four transitions are divided into two non-degenerate pairs of degenerate transitions.
The valley and spin breakdown of the intraband absorption peak, presented in Fig.~\ref{fig:chixx_doped_+schemes}-(b), shows that the degenerate transitions yield different but comparable intensities.
In addition, it also shows that the non-degenerate transitions cannot be resolved in energy.
This is explained by the presence of a broadening parameter, $\Gamma=7\si{\milli\electronvolt}$, which blurs the small energy splitting between the peaks.

The broadening parameter also makes the intraband optical spectrum robust with respect to variations in the temperature.
The small temperature dependency can be understood with the help of the scheme in Fig.~\ref{fig:chixx_doped_+schemes}-(d).
Looking at the short arrows ---which represent intraband transitions that respect the optical selection rules---, we see that the green one marks the only allowed transition at $T=0$.
At finite temperatures, other LLs are thermally activated (blue region) and enable more transitions (yellow arrows).
The absence of a noticeable temperature dependency is then obtained because, up to the first Pauli blocked transitions (red arrows), the variation in energy of these transitions is small compared to $\Gamma$.
This is a consequence of the highly linear dispersion of the LLs with $n$ in the regime $\Delta_{\tau s}^2 \gg \frac{1}{2} \left( \hbar \omega_0 \right)^2 n$.

Doping introduces new features in the interband contributions to $\chi_{xx} (\omega)$. 
First, we observe a blue shift of the absorption threshold, associated with the filling of LLs in the conduction band, for the case of a n-doped system, or the depletion of LLs in the valence band, in the case of p-doping.  
Second, we obtain a lineshape that carries a significant temperature dependency, as seen in Fig.~\ref{fig:chixx_doped_+schemes}-(a).
At $T=0$, the lineshape features a similar double step structure that reflects the strong SOC.
However, at room temperature, this feature is smoothed out and the explanation is self-evident in the scheme of Fig.~\ref{fig:chixx_doped_+schemes}-(d).
Looking at the long arrows, which mark the less energetic interband transitions in play due to thermal activation (within the same color code as before), it is clear that, in contrast with the intraband optical spectrum, the increase of the temperature induces transitions that can be resolved in energy, which in turn leads to the disappearance of a clear double step structure.
It must be noted that our analysis does not include the reduction of the band gap with the increase of the temperature, expected due to thermal expansion of the lattice that widens the bands~\cite{Ashcroft1976}.

The most intriguing difference between the doped and undoped interband optical spectra is observed in the limit of $T=0$, whose validity is discussed below.
In the doped case, the height of the lowest energy interband peak in absorption is half of the others within the SOC plateau.
The origin of this ``half peak'' is explained through the Pauli exclusion principle. 
For a given $\tau s$, and since 0LLs are not in play, there are in general four degenerate interband transitions contributing to the absorption peaks, as depicted in Table~\ref{tab:transitions}.
However, for the half peak, two out of the four transitions are Pauli blocked, leading to a reduction of the intensity by half.
In Fig.~\ref{fig:chixx_doped_+schemes}-(d), the two blocked transitions are represented by the yellow dashed arrow, while the two allowed ones are represented by the long green arrow~\footnote{For the sake of clarity, we underline that each arrow in Fig.~\ref{fig:chixx_doped_+schemes}-(d) represents two transitions, as there are two possible combinations of $\tau$ and $s$ that yield $\tau s = +1$.}.
In practice, the limit $T=0$ is valid as long as the thermal activation does not change considerable the occupation of the LLs that are immediately above or below the Fermi level.
This is realized for $T \lesssim 0.5 B_0[\si{\tesla}] \si{\kelvin}$.

Interestingly, the elimination of two out of four transitions that results in the half peak also provides a way to induce both a valley and spin imbalance in TMDs using linearly polarized light.
The intensity of the four degenerate transitions is controlled by the matrix elements, in such a way that there are two equally strong and two equally weak oscillator strengths, as previously mentioned in Section~\ref{subsection:longitudinal_undoped}.   
For instance, Eq.~\eqref{eq:dipoles_ident} imposes that if some transition $\{n; \mathcal{V} \} \rightarrow \{n+1; \mathcal{C} \}$ is strong in the channel $\{\tau;s\}$, so it is the (counterpart) transition $\{n+1; \mathcal{V} \} \rightarrow \{n; \mathcal{C} \}$ in the channel $\{-\tau;-s\}$.
Now, in the case of the half peak, Pauli blocking occurs for transitions that are not counterpart of each other, which results on having only one of the two strong transitions active.
Therefore, the resulting absorption is overwhelmingly dominated by just one valley and one spin, as observed in Fig.~\ref{fig:chixx_doped_+schemes}-(c).
In fact, the intensities are so different that the contribution of the weak transition cannot be detected.

Our findings imply that driving a doped TMD with linearly polarized light can induce a nearly perfect spin and valley imbalance at some specific range of frequencies of the longitudinal magneto-optical absorption.
As we shall see in Section~\ref{section:Hall}, the same imbalance is also verified in the transverse response.
These findings permit to envision a mechanism for optical orientation and add value to the field of valleytronics.

\subsection{Doped system with a single Landau level polarized} 
\label{subsection:longitudinal_doped2}
We now briefly comment on the regime where the TMD is doped with electrons or holes up to the first 0LL in the conduction or valence band, respectively.
In this case, the system has a spin-polarized ground state.

It is straightforward to check that, at sufficiently low temperatures, a single valley and spin control can be achieved either at the intraband part of the longitudinal absorption spectrum or at the frequency of the less energetic transition in the interband part.
In this situation, the spin and valley selectiveness is not nearly perfect as a consequence of extremely unbalanced dipole matrix elements (as in Section~\ref{subsection:longitudinal_doped1}) but exact and based entirely on the optical selection rules.
This is strongly connected with the findings from Ref.~\onlinecite{Tabert2013}.

The carrier density needed to polarize a single LL is given by $|\rho| \simeq 2.4 \times 10^{10} B_0[\si{\tesla]}\si{\per \centi \meter \squared}$.  
Thus, the right combination of carrier density and magnetic field that leads to this regime seems within experimental reach. 

\section{Transverse susceptibility} 
\label{section:Hall}
In this section, we undertake the analysis of the dynamical transverse susceptibility, $\chi_{yx} (\omega)$, also known as Hall susceptibility.
As seen in Eq.~\eqref{eq:chi_cartesian_circular}, this quantity determines circular dichroism.
Therefore, it is relevant to model experiments that explore the magneto-optical Kerr effect and the Faraday rotation, for example.

At half filling, the contributions to $\chi_{yx} (\omega)$ coming from opposite valleys have opposite signs.
As a result, the total $\chi_{yx} (\omega)$ vanishes, although each valley yields a finite AC Hall response, as demonstrated in Appendix~\hyperref[sec:AppendixB]{B}. 
Thus, the application of an out-of-plane magnetic field ---which breaks time reversal symmetry--- is not sufficient to induce a Hall response in intrinsic TMDs.

For doped TMDs, the transverse susceptibility is no longer null and can be split into two terms, $\chi_{yx} (\omega) =  \chi^{\text{intra}}_{yx} (\omega) + \chi^{\text{inter}}_{yx} (\omega)$, which are determined by intra and interband types of optical transitions, respectively.
For simplicity, we take $T=0$ and consider a system in which the 0LLs cannot participate in the optical transitions.
This regime is realized for $T \lesssim 0.5 B_0[\si{\tesla}] \si{\kelvin}$ and $\mu > \text{max} \left( E^\eta_{1,\mathcal{C}} \right)$ or $\mu < \text{min} \left( E^\eta_{1,\mathcal{V}} \right)$.
Within these considerations, we obtain largely simplified analytical expressions for $\chi^{\text{intra}}_{yx} (\omega)$ and $\chi^{\text{inter}}_{yx} (\omega)$, given by
\begin{equation}
\text{\small $
\chi^{\text{intra}}_{yx} (\omega) = 
\mathrm{i} \frac{\text{sign}(\mu) \left( \hbar \omega + \mathrm{i} \Gamma \right)}{\pi l_B^2 \varepsilon_0} \sum_\eta 
\frac{\left| {d_x^{\eta}}_{\{n_\mu;\text{sign}(\mu)\}}^{\{n_\mu + 1;\text{sign}(\mu)\}} \right|^2}
{\left( \hbar \omega^\eta_{n_\mu} \right)^2 - \left( \hbar \omega + \mathrm{i} \Gamma \right)^2},
$} 
\label{eq:chi_xy_intra}
\end{equation}
\begin{equation}
\text{\small $
\chi^{\text{inter}}_{yx} (\omega) = 
\mathrm{i} \frac{\text{sign}(\mu) \left( \hbar \omega + \mathrm{i} \Gamma \right)}{\pi l_B^2 \varepsilon_0} \sum_\eta 
\frac{\left| {d_x^{\eta}}_{\{n_\mu;-\text{sign}(\mu)\}}^{\{n_\mu + 1;\text{sign}(\mu)\}} \right|^2}
{\left( \hbar \Omega^\eta_{n_\mu} \right)^2 - \left( \hbar \omega + \mathrm{i} \Gamma \right)^2},
$} 
\label{eq:chi_xy_inter1}
\end{equation}
where $n_\mu = n_F - \frac{1-\text{sign}(\mu)}{2}$ is introduced for convenience and corresponds to the last occupied LL if $\mu>0$ or to the first empty one if $\mu < 0$, while 
\begin{equation}
\hbar \omega^\eta_{n_\mu} = \left| E^\eta_{n_\mu + 1,\text{sign}(\mu)} - E^\eta_{n_\mu,\text{sign}(\mu)} \right| 
\end{equation}
and
\begin{equation}
\hbar \Omega^\eta_{n_\mu} = \left| E^\eta_{n_\mu + 1,\text{sign}(\mu)} - E^\eta_{n_\mu,-\text{sign}(\mu)} \right|
\end{equation}
are the energies of the intraband and interband transitions contributing to the Hall response, respectively.
For clarity purposes, we note that the sum over LLs, present in the general expression for $\chi_{yx} (\omega)$, is taken care of by the fact that all the (canceling) contributions that lead to a null AC Hall response in the undoped regime can be removed. 

\begin{figure*}
 \includegraphics[width=2\columnwidth]{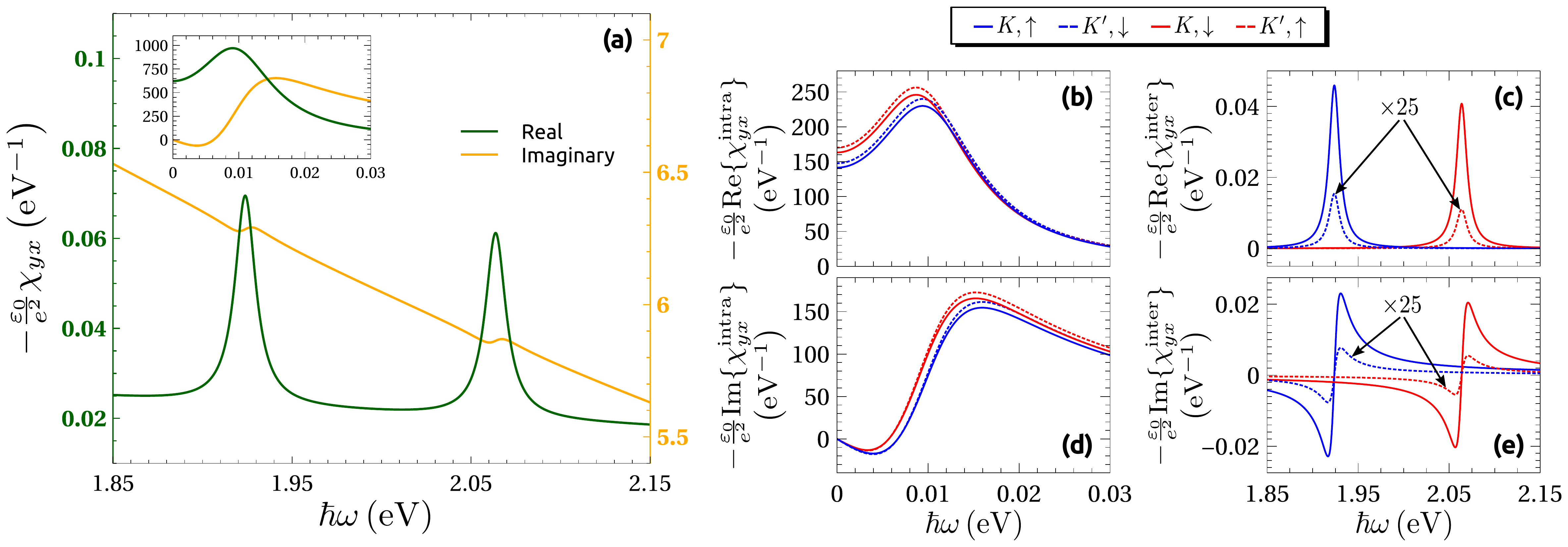}
 \caption{
 (Color online) (a) Hall susceptibility, $\chi_{yx}$, as a function of the photon energy, in a doped (Fermi level $\mu = 1\si{\electronvolt}$) monolayer \ch{MoS2} at zero absolute temperature and for a magnetic field of $50\si{\tesla}$.
 (b-e) Valley and spin breakdown of the real (b,c) and imaginary (d,e) parts of (a), divided in the (non-canceling) contributions that come from intraband (b,d) and interband (c,e) optical transitions.
  The valley and spin breakdown of the interband optical spectrum reveals a dominant contribution of transitions within the $K$ valley.}
 \label{fig:chixy_doped}
\end{figure*}

In Fig.~\ref{fig:chixy_doped}, we present typical results in the regime for which Eqs.~\eqref{eq:chi_xy_intra} and \eqref{eq:chi_xy_inter1} are valid.
The doping case is the same as the one considered in Section~\ref{subsection:longitudinal_doped1}.
Additionally, the choice of the parameters allows for a direct comparison of these results with the ones obtained in Fig.~\ref{fig:chixx_doped_+schemes}.

In contrast to the longitudinal response, resonance peaks are observed in the real part of the Hall susceptibility.
This is justified by the fact that absorption is described by the susceptibility tensor in its diagonal form, Eq.~\eqref{eq:chi_cartesian_circular}, i.e., in the circular basis.
In this basis, the contribution to the imaginary part of $\chi_\pm (\omega)$ comes from the real part of $\chi_{yx} (\omega)$.
Analytically, this is also verified through Eq.~\eqref{eq:chi_1} by the presence of an extra overall imaginary unit when comparing the expressions for $\chi_{xx} (\omega)$ and $\chi_{yx} (\omega)$.

The results shown in Fig.~\ref{fig:chixy_doped}-(a) imply genuine (as opposed to valley-resolved) circular dichroism.
Through Eq.~\eqref{eq:chi_cartesian_circular}, we see that $\text{Re} \{ \chi_{yx} \} \neq 0$ leads to a differential absorption of $\sigma^+$ and $\sigma^-$ photons.
This effect is stronger at the resonant frequencies.

The spin and valley breakdown of the Hall response, shown in Figs.~\ref{fig:chixy_doped}-(b-e), reveals that interband absorption is dominated by the $K$ valley.
Due to SOC, this also implies a spin imbalance, given that transition energies are related by spin-valley coupling.
The origin of this result is completely analogous to the discussion of the half peak in Section~\ref{subsection:longitudinal_doped1}.

As in Section~\ref{subsection:longitudinal_doped2}, it is straightforward to verify that, at sufficiently low temperatures, a TMD with a single LL polarized induces a (perfect) spin and valley imbalance in the Hall response, which is based entirely on the optical selection rules.
Evidently, the transitions responsible for this phenomenon involve the 0LLs.

\section{Response to circularly polarized light} 
\label{section:circular}
\begin{figure}
 \includegraphics[width=\columnwidth]{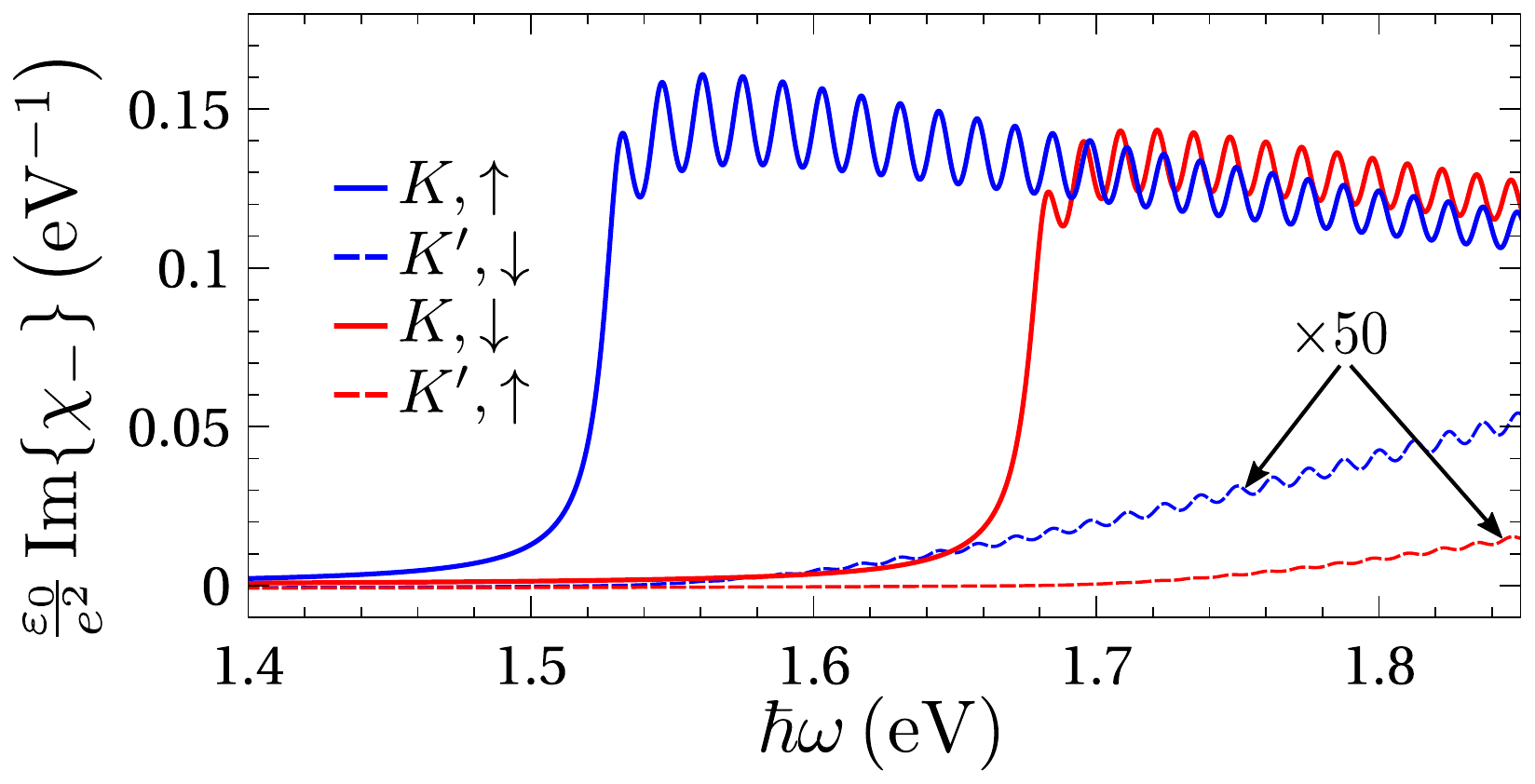}
 \caption{(Color online) 
 Imaginary part of the susceptibility to left-handed circularly polarized light, $\chi_-$, as a function of the photon energy, in monolayer \ch{MoS2} and for a magnetic field of $30\si{\tesla}$ (results independent of the temperature and resolved in the valley and spin contributions).
 The peaks in $\text{Im} \{ \chi_- \}$, which are directly related with absorption of left-handed photons, reveal a valley-selective circular dichroism towards the $K$ valley.
 Results for right polarization are the same with opposite spin and valley.}
 \label{fig:chipm_mugap}
\end{figure}

The thorough study of $\chi_{xx}$ and $\chi_{yx}$ presented in the last two sections permits to address the magneto-optical response of TMDs to circularly polarized light.
Here, we focus on the absorptive part of $\chi_\pm (\omega) = \chi_{xx} (\omega) \pm \mathrm{i} \chi_{yx} (\omega)$ at half filling.
In Fig.~\ref{fig:chipm_mugap}, we show representative results, obtained for undoped \ch{MoS2} and $B_0 = 30 \si{\tesla}$.
The analysis follows below.

It is apparent that the absorption of $\sigma^-$ ($\sigma^+$) photons is dominated by the $K$ ($K'$) valley. 
Thus, the well-known~\cite{Yao2008,Ezawa2013,Xu2014} valley-resolved circular dichroism at $B_0 = 0$ is preserved at finite field.
Given that $\chi_{xx} (\omega)$  has equal contributions from both valleys, the valley imbalance is fully controlled by $\chi_{yx} (\omega)$.
This is made possible by the fact that, in the intrinsic case, $\chi_{yx} (\omega)$ is non-zero for each valley, even though the sum over valleys yields a vanishing AC Hall response.

To gain insight about the origin of the valley-selective circular dichroism, we make the limit of no impurities, $\Gamma \rightarrow 0^+$, and use the Sokhotski-Plemelj theorem to write
\begin{equation}
\begin{split}
& \text{Im} \{ \chi_\pm (\omega) \} = \pm \sum_\eta \sum_{\{n;\lambda\}}
\frac{\lambda}{l_B^2 \varepsilon_0} \left| {d_x^{\eta}}_{\{n;\lambda\}}^{\{n+1;-\lambda\}} \right|^2 \times \\
& \quad \times
\delta \left( E_{n,\lambda}^\eta - E_{n+1,-\lambda}^\eta \mp \hbar \omega \right), 
\label{eq:chi_pm_limit}
\end{split}
\end{equation}
where we have also used that, in the undoped regime,
\begin{equation}
f\left(E_{n+1,\lambda'}^\eta\right) - f\left(E_{n,\lambda}^\eta\right) = \lambda \delta_{\lambda',-\lambda}.
\end{equation}
Looking at Eq.~\eqref{eq:chi_pm_limit}, we observe that the Dirac Delta implies $\lambda=\mathcal{C}/\mathcal{V}$ for right/left polarization.
This relation blocks counterpart transitions for the whole interband optical spectrum, in the same way that doping blocks a specific set of counterpart interband transitions that contribute to $\chi_{xx}$ and $\chi_{yx}$.
As a result, we get highly unbalanced valley contributions at any $\omega$ of the interband absorption, which are determined exclusively by the magnitude of the dipole matrix elements.

The presence of SOC interactions is only reflected by the splitting of the lineshapes that correspond to different spin contributions within the same valley.
Thus, the valley-selective circular dichroism is independent of SOC and only determined by the $\tau$ dependency of the kinetic term in the Hamiltonian.

These results show that the well-established optically-induced valley polarization for intrinsic TMDs~\cite{Cao2012} remains upon application of an out-of-plane magnetic field.
In addition, the analytical approach to this problem unveils that the valley-selective circular dichroism is not a selection rule that completely cancels absorption in one valley, but a consequence of extremely unbalanced dipole matrix elements.

\section{Exchange self-energy corrections} 
\label{section:exchange}
We now turn our attention to how the electronic and optical properties discussed before are modified due to Coulomb interactions.
In particular, we keep track of corrections up to the self-energy (SE) level, which lead to the renormalization of the electronic band structure and thus affect the optical response by changing the frequency of the transitions in play.
Since the dipole matrix elements remain identical, the main features of the magneto-optical response of TMD monolayers are maintained at this level of approximation.
The inclusion of these effects is carried out within the same EOM formalism.

\subsection{Keldysh potential}
In order to account for electron-electron repulsions in a 2D landscape, we replace the typical Coulomb potential by the Keldysh potential~\cite{Cudazzo2011}.
In the direct space, the Keldysh energy potential between two electrons in $\bm{r}$ and $\bm{r}'$, $U(\bm{r}-\bm{r}')$, has a rather intricate form. 
In contrast, its Fourier transform yields a more transparent expression, given by
\begin{equation}
U(\bm{q}) = \frac{e^2}{2 \varepsilon_0} \frac{1}{q \left( r_0 q + 1 \right)}, 
\label{eq:Keldysh}
\end{equation}
where $\bm{q} = (q_x,q_y)$ is the transferred momentum and $r_0$ is a material-dependent constant that measures the deviation from the 2D Coulomb energy potential, which is recovered making $r_0 = 0$.

When in presence of a dielectric medium with relative permittivity $\varepsilon_r$, Eq.~\eqref{eq:Keldysh} is modified by the transformation $r_0 q + 1 \rightarrow r_0 q + \varepsilon_r$.
For simplicity, we assume TMDs in vacuum or suspended in air ($\varepsilon_r \simeq 1$), thus ignoring screening effects due to the presence of dielectric media.
The magnitude of the band renormalization so obtained is therefore an upper limit.

\subsection{Exchange self-energy: analytical expressions}
Disregarding coupling between different valleys, we write the (two-particle) Hamiltonian that accounts for electron-electron interactions as
\begin{equation}
\hat{H}_{ee} (t) = \frac{1}{2} \sum_{\substack{\tau \\ s, s'}} \sum_{\substack{\alpha_1, \alpha_2 \\ \alpha_3, \alpha_4}} U^{\tau,s,s'}_{\substack{\alpha_1, \alpha_2 \\ \alpha_3, \alpha_4}}
\hat{c}^\dagger_{\alpha_1,\tau,s} \hat{c}^\dagger_{\alpha_2,\tau,s'} \hat{c}_{\alpha_3,\tau,s'} \hat{c}_{\alpha_4,\tau,s}, 
\label{eq:Hee}
\end{equation}
where
\begin{equation}
U^{\tau,s,s'}_{\substack{\alpha_1, \alpha_2 \\ \alpha_3, \alpha_4}} = 
\int_{\mathbb{R}^2} \frac{d\bm{q}}{(2\pi)^2} \ U(q) 
F^{\tau,s}_{\alpha_1,\alpha_4}(\bm{q}) F^{\tau,s'}_{\alpha_2,\alpha_3}(-\bm{q}) 
\label{eq:Coulomb_integrals}
\end{equation}
are the Coulomb integrals and 
\begin{equation}
F^{\tau,s}_{\alpha,\alpha'}(\bm{q}) = \int_A d\bm{r} \ 
\mathrm{e}^{\mathrm{i} \bm{q} \cdot \bm{r}} 
\left[ \psi^{\tau,s}_\alpha(\bm{r}) \right]^\dagger \psi^{\tau,s}_{\alpha'}(\bm{r}) 
\label{eq:structure_factors}
\end{equation}
the structure factors.
In Eq.~\eqref{eq:Hee}, the time dependency of the fermionic operators is omitted to shorten notation.
The exclusion of inter-valley contributions is justified by the large momentum difference between $K$ and $K'$, which implies a large transferred momentum that in turn suppresses $U(q)$ and consequently the inter-valley Coulomb integrals.

The following task is to include $\hat{H}_{ee} (t)$ in the total Hamiltonian, Eq.~\eqref{eq:Htotal}, and obtain the new (interacting) EOM.
This task boils down to the calculation of the commutator $\left[\hat{H}_{ee}(t), \hat{T}^\eta_{\alpha,\alpha'}(t) \right]$, whose result is shown in Appendix~\hyperref[subsec:AppendixC1]{C.1}.
Among the new terms, we then identify and keep the ones that lead to a band renormalization.
Random phase approximation and linear response regime are implied in this last step and the details regarding this manipulation can be found in Appendix~\hyperref[subsec:AppendixC2]{C.2}.
As final result, we find that the energy bands are renormalized as 
\begin{equation}
\left( E^\eta_\alpha \right)_{\text{renorm}} = E^\eta_\alpha + \Sigma^\eta_\alpha,
\end{equation}
where 
\begin{equation}
\Sigma^\eta_\alpha = - \sum_{\alpha'} f\left( E^\eta_{\alpha'} \right) U^{\tau,s,s}_{\substack{\alpha', \alpha \\ \alpha', \alpha}} 
\label{eq:SE}
\end{equation}
are the exchange SE corrections.
As usual, we observe that the exchange corrections to energy bands with a given spin come from electrons in bands with the same spin.

The Coulomb integrals can be reduced to one-dimensional quadratures (see Appendix~\hyperref[subsec:AppendixC3]{C.3} for details).
At $T=0$, Eq.~\eqref{eq:SE} is simplified into
\begin{equation}
\Sigma^\eta_\alpha = - \sum_{\{n',\lambda'\} \in \text{occ.}} D^{\eta}_{\substack{\{n,\lambda\} \\ \{n',\lambda'\}}} I^{\eta}_{\substack{\{n,\lambda\} \\ \{n',\lambda'\}}}, 
\label{eq:SEsimp}
\end{equation}
where $D^{\eta}_{\substack{\{n,\lambda\} \\ \{n',\lambda'\}}}$ are real constants defined as
\begin{equation}
D^{\eta}_{\substack{\{n,\lambda\} \\ \{n',\lambda'\}}} = \frac{1}{2^{|n-n'|} } \left( C^\eta_{n,\lambda} C^\eta_{n',\lambda'} \right)^2,
\end{equation}
$I^{\eta}_{\substack{\{n,\lambda\} \\ \{n',\lambda'\}}}$ are integrals given by
\begin{equation}
\begin{split}
& \text{\small $
I^{\eta}_{\substack{\{n,\lambda\} \\ \{n',\lambda'\}}} = \frac{1}{l_B^2} \int_0^{+\infty}\frac{d\bar{q}}{2\pi} \ 
\bar{q}^{2|n-n'|+1} U\left( \frac{\bar{q}}{l_B} \right) \mathrm{e}^{-\bar{q}^2/2} \times
$} \\ 
& \text{\small $
\quad \times \Bigg| \tilde{L}^{|n-n'|}_{(n_\tau, n'_\tau)} \left( \frac{\bar{q}^2}{2} \right) + 
B^\eta_{n,\lambda} B^\eta_{n',\lambda'} \tilde{L}^{|n-n'|}_{(n_\tau + \tau, n'_\tau + \tau)} \left( \frac{\bar{q}^2}{2} \right) \Bigg|^2,
$} 
\label{eq:SEintegrals}
\end{split}
\end{equation}
and the notation $\{n',\lambda'\} \in \text{occ.}$ means that the sum runs over occupied states only. 
In Eq.~\eqref{eq:SEintegrals}, we have defined $\bar{q} \equiv l_B q$, $\tilde{L}^{|n-n'|}_{(b,c)} \equiv \sqrt{\frac{\text{min}(b,c)!}{\text{max}(b,c)!}} L^{|n-n'|}_{\text{min}(b,c)}$ for $\text{min}(b,c) \in \mathbb{N}^0$ ($L^{|n-n'|}_{\text{min}(b,c)}$ are the associated Laguerre polynomials) and $\tilde{L}^{|n-n'|}_{(b,c)} \equiv 0$ for $\text{min}(b,c)=-1$. 
Moreover, we remind that $n_\tau \equiv n  - \frac{1+\tau}{2}$.

In order to evaluate Eq.~\eqref{eq:SEsimp}, it is clear that a cutoff is again required, as the summation implied extends over an infinity of valence states.
Furthermore, we have verified numerically that the summation diverges logarithmically with the LL cutoff, $n_\text{cut}$.
Even when dealing with energy differences, this was checked to lead to corrections that are, to some extent, cutoff-dependent.
To fix $n_\text{cut}$, we start by counting the total number of electrons in a TMD sample of area $A$.
At half filling, we get $2 A/A_\text{u.c.}$, where $A_\text{u.c.} = \frac{\sqrt{3}}{2} a^2$ is the area of the hexagonal unit cell with lattice parameter $a \simeq 3.15\si{\angstrom}$~\cite{Ding2011}.
Then, this number is divided by $4$ (to account for spin and valley) and matched to the number of electronic states in $n_\text{cut}$ LLs.
Given the degeneracy of the LLs, $\frac{A}{2 \pi l_B^2}$, we obtain that
\begin{equation}
 n_\text{cut} = \frac{\pi l_B^2}{A_\text{u.c.}} \simeq \frac{24000}{B_0[\si{\tesla}]}
\end{equation}
is the number of filled LLs per spin and valley.

In the computations that follow, we use the material-dependent parameters listed in Table~\ref{tab:parametersxc}.

\begin{table}
 \begin{tabular}{l | c c c c c}
      & $\hbar v_F \left( \si{\electronvolt \angstrom} \right)$ 
      & $\Delta \left( \si{\electronvolt} \right)$ 
      & $\Delta_{\text{SOC}}^\mathcal{V} \left( \si{\electronvolt} \right)$
      & $\Delta_{\text{SOC}}^\mathcal{C} \left( \si{\electronvolt} \right)$ 
      & $r_0 \left( \si{\angstrom} \right)$ \\ \hline
      \ch{MoS2} & $3.51$ & $0.83$ & $0.148$ & $-0.003$ & $41.5$ \\
      \ch{WS2} & $4.38$ & $0.90$ & $0.430$ & $+0.029$ & $37.9$ \\
      \ch{MoSe2} & $3.11$ & $0.74$ & $0.184$ & $-0.021$ & $51.7$ \\
      \ch{WSe2} & $3.94$ & $0.80$ & $0.466$ & $+0.036$ & $45.1$ \\
 \end{tabular}
 \caption{List of parameters used in the numerical computation of the exchange self-energy corrections for different transition metal dichalcogenides.
 Values in the first and second, third and forth, and last columns were taken from Ref.~\onlinecite{Xiao2012}, Ref.~\onlinecite{Liu2013}, and Ref.~\onlinecite{Berkelbach2013}, respectively.}
 \label{tab:parametersxc}
\end{table}

\subsection{Renormalized optical transition energies}
As a direct application of the calculations presented above, we study how a selected set of optical transitions is renormalized in energy due to the exchange SE corrections, at $T=0$.
We consider different TMDs and focus on the following cases:
\begin{itemize}
  \item Fermi level in the gap. 
  Interband transitions: $\mathcal{T}^{K,s}_{ \{ 0,\mathcal{V} \} \rightarrow \{ 1,\mathcal{C} \} } \equiv E^{K,s}_{1,\mathcal{C}} - E^{K,s}_{0,\mathcal{V}}$ and 
  $\mathcal{T}^{K',s}_{ \{ 1,\mathcal{V} \} \rightarrow \{0,\mathcal{C}\} } \equiv E^{K',s}_{0,\mathcal{C}} - E^{K',s}_{1,\mathcal{V}}$.
  From the renormalization of these transition energies, we obtain the renormalized energy thresholds that define the SOC plateau observed in the absorption spectrum of intrinsic TMDs (see Figs.~\ref{fig:chixx_mugap} and \ref{fig:chipm_mugap}).
  Evidently, the exchange-corrected value of $\mathcal{T}^{K,\uparrow}_{\{ 0,\mathcal{V} \} \rightarrow \{ 1,\mathcal{C} \}} = \mathcal{T}^{K',\downarrow}_{\{ 1,\mathcal{V} \} \rightarrow \{0,\mathcal{C}\}}$ corresponds to the renormalized band gap.
 
  \item System doped with electrons or holes up to the first 0LL.
  Intraband transitions: $\mathcal{T}^{K,\uparrow}_{ \{ 1,\mathcal{V} \} \rightarrow \{ 0,\mathcal{V} \} } \equiv E^{K,\uparrow}_{0,\mathcal{V}} - E^{K,\uparrow}_{1,\mathcal{V}}$, for p-doping, and $\mathcal{T}^{K',*}_{ \{ 0,\mathcal{C} \} \rightarrow \{ 1,\mathcal{C} \} } \equiv E^{K',*}_{1,\mathcal{C}} - E^{K',*}_{0,\mathcal{C}}$, for n-doping, where $*=\uparrow$ if $\Delta^\mathcal{C}_{\text{SOC}} > 0$ and vice-versa.
  In this regime, these optical transitions lead to intraband peaks in the absorption spectrum that are spin- and valley-selective. 
\end{itemize}
In the undoped case, both the interband optical spectrum and the exchange SE corrections are independent of $T$.
For doped systems the limit $T=0$ is only valid as long as $T \lesssim 0.5 B_0[\si{\tesla}] \si{\kelvin}$ and provides an upper limit for the renormalization of the intraband transition energies.

In Table~\ref{tab:renormalization}, we present the results obtained for $B_0 = 10\si{\tesla}$.
These results show the usual tendency of the Hartree-Fock approximation to enhance energy gaps obtained through standard local density functional theory calculations. 
However, it must be noted  that, in optical spectroscopic measurements, absorption occurs for photon energies below the exchange-corrected values due to excitonic effects.

 \begin{table*}
  \begin{tabular}{l | c c c c}
      & $\mathcal{T}^{K,\uparrow}_{ \{ 0,\mathcal{V} \} \rightarrow \{ 1,\mathcal{C} \} } \left( \si{\electronvolt} \right)$
      & $\mathcal{T}^{K,\downarrow}_{ \{ 0,\mathcal{V} \} \rightarrow \{ 1,\mathcal{C} \} } \left( \si{\electronvolt} \right)$
      & $\mathcal{T}^{K,\uparrow}_{ \{ 1,\mathcal{V} \} \rightarrow \{ 0,\mathcal{V} \} } \left( \si{\milli\electronvolt} \right)$ 
      & $\mathcal{T}^{K',*}_{ \{ 0,\mathcal{C} \} \rightarrow \{ 1,\mathcal{C} \} } \left( \si{\milli\electronvolt} \right)$ \\ \hline
      \ch{MoS2} & $1.587,\ 2.454$ & $1.738,\ 2.717$ & $2.4,\ 105.4$ & $2.4,\ 103.5$ \\
      \ch{WS2} & $1.603,\ 2.567$ & $2.003,\ 3.024$ & $3.6,\ 107.9$ & $2.9,\ 105.1$ \\
      \ch{MoSe2} & $1.380,\ 2.202$ & $1.584,\ 2.433$ & $2.1,\ 101.2$ & $2.1,\ 100.7$ \\
      \ch{WSe2} & $1.388,\ 2.240$ & $1.818,\ 2.729$ & $3.4,\ 104.9$ & $2.6,\ 102.9$ \\
  \end{tabular}
\caption{Renormalization in energy of a selected set of optical transitions (described in the text) for different transition metal dichalcogenides and a magnetic field of $10\si{\tesla}$: 
bare and exchange-corrected values (computed at zero absolute temperature) separated by commas, in the respective order.
Results obtained for $\mathcal{T}^{K,s}_{ \{ 0,\mathcal{V} \} \rightarrow \{ 1,\mathcal{C} \} }$ are equal to the ones for $\mathcal{T}^{K',-s}_{\{ 1,\mathcal{V} \} \rightarrow \{0,\mathcal{C}\}}$.}
  \label{tab:renormalization}
\end{table*}

For intrinsic TMDs, we find a band gap correction whose magnitude is comparable to the renormalization of the direct band gap in the absence of external magnetic fields~\cite{Chaves2017}.
In the case of the intraband transitions between adjacent LLs, the exchange-corrected values obtained are most likely a severe overestimation of what should be observed in optical experiments.
In fact, Kohn's theorem~\cite{Kohn1961} states that the cyclotron resonance frequency of an electron gas is not altered by electron-electron interactions. 
Although this theorem ignores the coupling to the lattice~\cite{Ando1982}, far-infrared spectroscopy probing the cyclotron frequency of the 2D electron gas formed in silicon inversion layers~\cite{Jr.1974} has revealed a good agreement between the experiment and the independent-electron theory.
The applicability of Kohn's theorem for Dirac electrons has been discussed in the literature~\cite{Roldan2010}.

Kohn's theorem implies the existence of interaction-independent collective modes that are relevant for optical spectroscopic measurements. 
However, this theorem does not preclude that the quasiparticle spectrum, probed directly through other experiments, can be strongly renormalized by interactions. 
Thus, scanning tunneling microscopy (STM) or a combination of angle-resolved photoemission spectroscopy (ARPES) and inverse ARPES could be used to investigate the renormalization of the LL energies due to Coulomb interactions.

\subsection{Renormalization of the spin-orbit splitting}
We now discuss an exchange-driven mechanism to enhance the spin-orbit splitting.
Since the 0LLs do not disperse with the magnetic field, the energy difference between the two $n=0$ LLs in the conduction/valence band is given by $\Delta_{\text{SOC}}^{\mathcal{C}/\mathcal{V}}$.
As shown in Table~\ref{tab:parametersxc}, first-principle calculations predict values of $\Delta_{\text{SOC}}^{\mathcal{C}}$ relatively small compared to those of $\Delta_{\text{SOC}}^{\mathcal{V}}$.
These first-principle results were obtained for undoped TMDs, in the absence of external fields.
Here, we consider the renormalization of $\Delta_{\text{SOC}}^{\mathcal{C}}$, due to SE corrections, for doped systems and in the presence of an out-of-plane magnetic field.

We take as example the case of a monolayer \ch{MoSe2}, for which $\Delta_{\text{SOC}}^{\mathcal{C}} = -21\si{\milli\electronvolt}$ in the undoped regime.
At the Hartree-Fock level, it is clear that, in order to maximize the renormalization of this splitting, the Fermi level should lie between the two $n=0$ LLs in the conduction band.
For this matter, we consider the material doped with electrons up to the lowest energy 0LL.
In addition, the system should be cooled down such that there is no significant thermal activation of the unoccupied 0LL.
For the calculations, we take $T=0$, which is valid as long as $k_B T \ll \Delta_{\text{SOC}}^{\mathcal{C}}$.

In the regime described above, the energy of the unoccupied 0LL is renormalized due to valence states only.
On the other hand, the energy of the polarized 0LL is renormalized by states in the valence bands and in the 0LL itself.
When computing the difference, the dominant contribution comes from the auto SE correction, i.e., the exchange SE correction to the occupied 0LL due to itself.
The origin of the other contributions, which come from corrections due to the $n \neq 0$ LLs in the valence band that do not cancel each other, can be traced back to the presence of SOC interactions in the model.

\begin{figure}
 \includegraphics[width=\columnwidth]{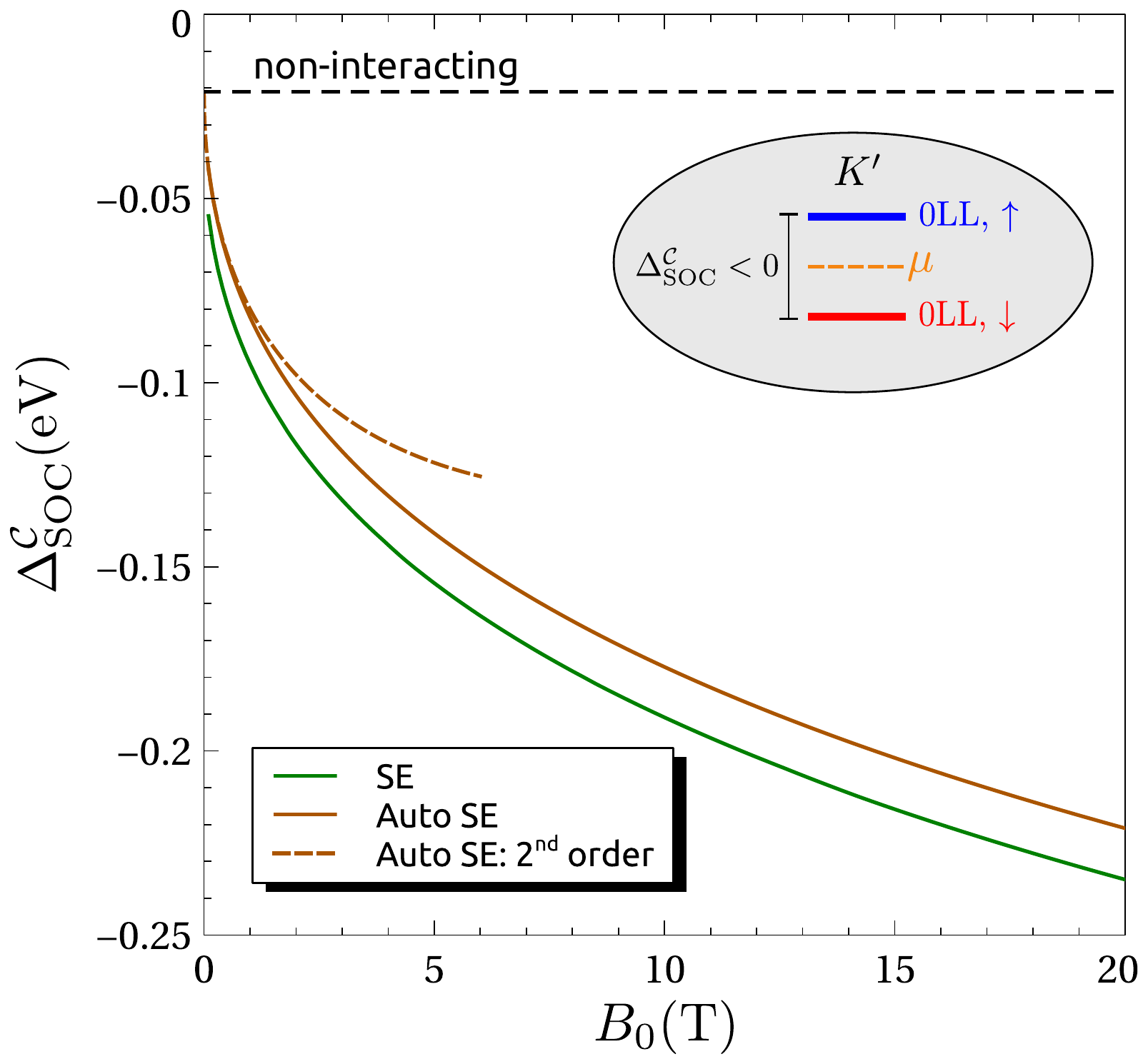}
 \caption{(Color online) Spin-orbit splitting of the zeroth Landau levels (0LLs) in the conduction band of \ch{MoSe2}, renormalized by the exchange self-energy (SE) corrections (computed at zero absolute temperature), as a function of the magnetic field.
 The Fermi level, $\mu$, is kept between the two spin split 0LLs in the conduction band, such that only the lowest energy 0LL in the $K'$ valley is polarized, as depicted in the cartoon.
 The horizontal black dashed line corresponds to the non-interacting reference, whereas the others correspond to exchange-corrected values that include the complete SE corrections (green solid line), the contribution of the auto SE only (brown solid line), and a low-field second order Taylor expansion of the former (brown dashed line).
 These results reveal a large exchange-driven enhancement of the splitting, which increases with the intensity of the magnetic field and approaches the non-interacting value in the limit of zero field.}
 \label{fig:SOCc_renorm}
\end{figure}

In Fig.~\ref{fig:SOCc_renorm}, we plot the evolution of the renormalized spin-orbit splitting of the 0LLs in the conduction band of \ch{MoSe2}, as a function of the magnetic field.
We present results that include the complete SE corrections, the contribution of the auto SE only, and a low-field approximation of the former (see derivations below).
The carrier density implied to keep only the lowest energy 0LL polarized is $\rho \simeq -2.4 \times 10^{10} B_0[\si{\tesla]}\si{\per \centi \meter \squared}$.
The analytical expression for the auto SE correction reads
\begin{align}
& \tilde{\Sigma}^\eta_\text{0LL} = - D^{\eta}_{\substack{\text{0LL} \\ \text{0LL}}} I^{\eta}_{\substack{\text{0LL} \\ \text{0LL}}} \nonumber
\\ & \phantom{\tilde{\Sigma}^\eta_\text{0LL}}
= -\frac{e^2}{4 \pi \varepsilon_0} \frac{1}{l_B} \int_0^{+\infty} d\bar{q} \ \frac{ \mathrm{e}^{-\bar{q}^2/2} }{ \frac{r_0}{l_B} \bar{q} + 1 } \nonumber
\\ & \phantom{\tilde{\Sigma}^\eta_\text{0LL}}
= -\frac{e^2}{4 \pi \varepsilon_0} \frac{ \mathrm{e}^{-\frac{l_B^2}{2 r_0^2 }} }{2 r_0} \left[ \frac{\pi}{\mathrm{i}} \text{erf} \left(\mathrm{i} \frac{l_B}{\sqrt{2} r_0} \right) - \text{Ei} \left( \frac{l_B^2}{2 r_0^2} \right) \right], 
\label{eq:autoSE}
\end{align}
where erf is the error function and Ei the exponential integral function.
In the limit of small $B_0$, Eq.~\eqref{eq:autoSE} can be simplified making a Taylor expansion around $\frac{r_0}{l_B} = 0$ which, up to second order, yields
\begin{equation}
\tilde{\Sigma}^\eta_\text{0LL} \simeq -\frac{e^2}{4 \pi \varepsilon_0} \frac{1}{r_0} \left( \frac{\sqrt{2 \pi}}{2}\frac{r_0}{l_B} - \frac{r_0^2}{l_B^2} \right). 
\label{eq:autoSE_Taylor}
\end{equation}
The validity of Eq.~\eqref{eq:autoSE_Taylor} is controlled by the ratio $\frac{r_0}{l_B}$, that scales as $0.2 \sqrt{B_0 [\si{\tesla}]}$ for \ch{MoSe2}.

The complete SE results show a large exchange-driven enhancement of $\Delta_{\text{SOC}}^{\mathcal{C}}$: even at a moderate field of $2 \si{\tesla}$, we obtain a renormalization in the order of $100\si{\milli\electronvolt}$.
It is also apparent that the exchange corrections are dominated by the auto SE contribution.
Thus, it becomes clear why the spin-orbit splitting increases with the intensity of the magnetic field: as $B_0$ ramps up, so it does the density of electrons in the occupied 0LL and therefore the magnitude of the renormalization.
Expectedly, we also observe that the exchange-corrected values approach the non-interacting reference as we decrease the intensity of the magnetic field.
This is verified analytically through Eq.~\eqref{eq:autoSE_Taylor} by noticing the absence of zeroth order terms in the low-field Taylor expansion of the auto SE correction.

The predictions of the Hartree-Fock calculations have to be contrasted with Larmor's theorem for spin-flip collective modes, excited with a zero wave vector perturbation~\cite{Dobers1988}. 
Analogously to Kohn's theorem, this theorem states that electron-electron interactions do not renormalize the energy of the $q=0$ spin-flip excitations, which must be equal to $g \mu_B B_0$.  
However, the theorem only holds for systems where the total spin is conserved, which is clearly not the case for TMDs, on account of the strong SOC interactions. 
On the other hand, vertex corrections are likely to reduce the large spin-flip energies predicted at the Hartree-Fock level~\cite{Mahan2013}.
In any case, experiments that probe the quasiparticle spectrum, such as STM and ARPES, might be able to capture the large shifts predicted by our calculations.

\section{Discussion and conclusions} 
\label{section:conclusion}
We have provided a thorough theoretical study of the optical properties of semiconducting TMD monolayers, described within the massive Dirac model, under the influence of strong out-of-plane magnetic fields that quantize the energy spectrum into a set of LLs.
We have analyzed in detail the longitudinal and transverse optical response, in both doped and undoped regimes, paying attention to the breakdown of the contributions coming from different spins and valleys.
We have also addressed the role of electron-electron interactions, treated at the Hartree-Fock level.

\subsection{Limits of the model}
Here, we briefly discuss some limitations of the model Hamiltonian applied in this work.
First, atomistic calculations~\cite{Chu2014,Lado2016} show a valley symmetry breaking of the LL spectrum that is not captured through Dirac models.
Thus, the resulting magneto-optical spectra should feature a valley splitting of the peaks.
Second, we have ignored the paramagnetic shift of the valence bands associated to the coupling between the magnetic field and the valley-dependent atomic orbital momentum, $L_z = \tau 2$, of the highest energy valence states~\cite{Kosmider2013}.
This results in another valley-dependent contribution.
Third, we have also ignored Zeeman splitting, that can be easily added to our results. 
Finally, we have not considered excitonic effects, that are expected to have a strong impact in the optical response. 
These are the scope of an incoming publication~\cite{Have2018}.
At charge neutrality, the excitonic effects not considered in this work are known to renormalize strongly the optical response functions. 
Therefore, our results in the undoped regime are meant to be taken, at most, as a qualitative description.
However, in the doped case, we expect our analysis to be robust against exciton formation.
To sustain this statement, we first note that the exciton size in  TMDs monolayers are not strongly affected by the presence of an out-of-plane magnetic field~\cite{Have2018}.
Then, we compare the $\bm{B}=\bm{0}$ exciton size ---typically in the order of a few nanometers~\cite{Chaves2017}--- with the 2D Thomas-Fermi screening length, which we have estimated to be $\sim 0.17\si{\nano\meter}$ and independent of the carrier density.  
These numbers lead us to conclude that excitons in TMDs are effectively screened in any doped regime for which the Thomas-Fermi approximation holds. 

\subsection{Main results}
We now summarize our main results.
At $\bm{B}=\bm{0}$, TMDs are known to present valley-dependent circular dichroism~\cite{Cao2012}: photons with a given circular polarization induce transitions in a valley-selective manner.
This permits to induce optical valley orientation.
Given that TMDs have strong SOC interactions, valley orientation also implies spin orientation in these materials.  
In this work, we have found that the application of an out-of-plane magnetic field preserves these effects, although the resulting optical spectrum contains a much richer structure.

In the case of doped TMDs, the application of the magnetic field brings two main novelties that are absent in the undoped regime:
\begin{enumerate}
\item The lowest energy peak in $\chi_{xx}(\omega)$ has dominant contributions from optical transitions within a single spin and valley (see Fig.~\ref{fig:chixx_doped_+schemes}-(c)).
As a result, at that energy, linearly polarized light can induce both a valley and spin imbalance. 
This provides a new mechanism for optical orientation, attained with linearly polarized light.

\item The AC Hall response is finite, as shown in Fig.~\ref{fig:chixy_doped}-(a).
This implies a net circular dichroism, i.e., a net difference in absorption of $\sigma^+$ and $\sigma^-$ photons.
\end{enumerate}

The main consequences of the exchange SE interactions are:
\begin{enumerate}
\item In the intrinsic case, the effective band gap is severely renormalized, resulting in a larger value.
\item In n-doped systems with a spin-polarized groundstate, our calculations show a strong exchange-driven renormalization of the spin-orbit splitting of the 0LLs in the conduction band, which exceeds $100\si{\milli\electronvolt}$ for $B_0 = 2\si{\tesla}$.   
\end{enumerate}
These results point out the strong influence of electron-electron interactions in the electronic and optical properties of doped TMDs.
Future work will address spin and valley Stoner instabilities driven by Coulomb interactions in doped TMDs (see for instance Ref.~\onlinecite{Szulakowska2018}).

\section*{Acknowledgements}
We thank Andre J. Chaves and Luis Brey for fruitful discussions.
G. C. thanks Departamento de F\'{i}sica Aplicada at Universidad de Alicante for their hospitality.
G. C. and J. F.-R. acknowledge financial support from FCT for the P2020-PTDC/FIS-NAN/4662/2014 project.
J. H. acknowledges financial support by the QUSCOPE Center, sponsored by the Villum foundation.
J. F.-R. acknowledges financial support from FCT for the P2020-PTDC/FIS-NAN/3668/2014 and the UTAP-EXPL/NTec/0046/2017 projects, as well as Generalitat Valenciana funding Prometeo2017/139 and MINECO-Spain (Grant No. MAT2016-78625-C2).
N. M. R. P. acknowledges financial support from the European Commission through the project ``Graphene-Driven Revolutions in ICT and Beyond'' (Ref. No. 785219) and the Portuguese Foundation for Science and Technology (FCT) in the framework of the Strategic Financing UID/FIS/04650/2013.
Additionally, N. M. R. P. acknowledges COMPETE2020, PORTUGAL2020, FEDER and the Portuguese Foundation for Science and Technology (FCT) for the Grants No. PTDC/FIS-NAN/3668/2013 and No. POCI-01-0145-FEDER-028114.

\appendix
\begin{widetext}

\section{Solution for the non-interacting equation of motion} 
\label{sec:AppendixA}
Due to the optical selection rules, the solution for the time evolution of the polarization density operator can be broken down into the problem of solving the EOM of a specific set of general operators $\hat{T}_{\alpha,\alpha'}^\eta (t) \equiv \hat{c}^\dagger_{\alpha',\eta}(t) \hat{c}_{\alpha,\eta}(t)$.
Introducing the notation
\begin{equation}
 \hat{c}_{\alpha,\eta}(t) \equiv 
 \begin{cases}
\hat{\mathcal{C}}_n &, n\geq1 \wedge \lambda=\mathcal{C}\\
\hat{\mathcal{V}}_n &, n\geq1 \wedge \lambda=\mathcal{V}\\
\hat{a}_0 &, \{n;\lambda\}=\text{0LL}
  \end{cases},
\end{equation}
where the dependency on $t$, $k_y$ and $\eta$ is omitted to compress notation~\footnote{There is no loss of generality in omitting $k_y$ and $\eta$ because the optical selection rules imply transitions that couple the same wave vector, valley and spin.}, the relevant set of pair of operators reads:
\begin{itemize}
 \item[1.] $\hat{a}^\dagger_0 \hat{\mathcal{C}}_1$, $\hat{a}^\dagger_0 \hat{\mathcal{V}}_1$ and hermitian conjugates, for transitions that involve the 0LLs.
 
 \item[2.] $\hat{\mathcal{C}}^\dagger_n \hat{\mathcal{V}}_{n+1}$, $\hat{\mathcal{C}}^\dagger_{n+1} \hat{\mathcal{V}}_n$ and the hermitian conjugates, for interband transitions between $n \neq 0$ LLs.
 
 \item[3.] $\hat{\mathcal{C}}^\dagger_n \hat{\mathcal{C}}_{n+1}$, $\hat{\mathcal{V}}^\dagger_n \hat{\mathcal{V}}_{n+1}$ and the hermitian conjugates, for intraband transitions between $n \neq 0$ LLs.
\end{itemize}
In what follows, we will keep track of only one of these pairs, $\hat{a}^\dagger_0 \hat{\mathcal{C}}_1$.
The derivation for the others follows straightforwardly and the final result is trivial to generalize, as we mention below.

After some straightforward algebra, the EOM for $\hat{a}^\dagger_0 \hat{\mathcal{C}}_1$ yields
\begin{equation}
 \frac{\hbar}{\mathrm{i}} \frac{d}{dt} \left( \hat{a}^\dagger_0 \hat{\mathcal{C}}_1 \right) = \left[ \hat{H}_0(t), \hat{a}^\dagger_0 \hat{\mathcal{C}}_1 \right] + 
 \left[ \hat{H}_I(t), \hat{a}^\dagger_0 \hat{\mathcal{C}}_1 \right],
 \label{eq:EOMa0c1}
\end{equation}
where
\begin{equation}
 \left[ \hat{H}_0(t), \hat{a}^\dagger_0 \hat{\mathcal{C}}_1 \right] = \left( E^\eta_{\text{0LL}} - E^\eta_{1,\mathcal{C}} \right) \hat{a}^\dagger_0 \hat{\mathcal{C}}_1
\end{equation}
and 
\begin{equation}
  \left[ \hat{H}_I(t), \hat{a}^\dagger_0 \hat{\mathcal{C}}_1 \right] = -\bm{\mathcal{E}}(t) \cdot \bigg[ {\bm{d}^{\eta}}_{\text{0LL}}^{\{1;\mathcal{C}\}} 
  \left( \hat{\mathcal{C}}^\dagger_1 \hat{\mathcal{C}}_1 - \hat{a}^\dagger_0 \hat{a}_0 \right) 
  + {\bm{d}^{\eta}}_{\text{0LL}}^{\{1;\mathcal{V}\}} \hat{\mathcal{V}}^\dagger_1 \hat{\mathcal{C}}_1 
  - {\bm{d}^{\eta}}_{\{2;\mathcal{V}\}}^{\{1;\mathcal{C}\}} \hat{a}^\dagger_0 \hat{\mathcal{V}}_2 
  - {\bm{d}^{\eta}}_{\{2;\mathcal{C}\}}^{\{1;\mathcal{C}\}} \hat{a}^\dagger_0 \hat{\mathcal{C}}_2 \bigg].
\end{equation}

In order to simplify the previous EOM, we start by taking its average with respect to the unperturbed Hamiltonian, $\hat{H}_0 (t)$, and then approximate $\braket{\hat{c}^\dagger_{\alpha,\eta}(t) \hat{c}_{\alpha,\eta}(t)}_0 \simeq \braket{\hat{c}^\dagger_{\alpha,\eta} \hat{c}_{\alpha,\eta}}_0$, where $\hat{c}^\dagger_{\alpha,\eta} / \hat{c}_{\alpha,\eta}$ are the creation/annihilation fermionic operators in the Schr\"{o}dinger representation.
The first simplification occurs because the expectation value of the time-independent number operator yields the Fermi-Dirac distribution, 
\begin{equation}
 \braket{\hat{c}^\dagger_{\alpha,\eta} \hat{c}_{\alpha,\eta}}_0 = f \left( E^\eta_{n,\lambda} \right) = \frac{1}{\mathrm{e}^{\beta \left( E^\eta_{n,\lambda} - \mu \right)} + 1},
\end{equation}
where $\mu$ is the Fermi level and $\beta \equiv 1/\left( k_B T \right)$ ($k_B$ is the Boltzmann constant and $T$ the absolute temperature).
In addition to that, we use the fact that the average value of the terms which connect either a) the same $n$ but different $\lambda$, or b) LL indexes that differ from $\pm 2$, is null.
This leads to
\begin{equation}
 \frac{\hbar}{\mathrm{i}} \frac{d}{dt} \braket{\hat{a}^\dagger_0 \hat{\mathcal{C}}_1}_0  = \left( E^\eta_{\text{0LL}} - E^\eta_{1,\mathcal{C}} \right) 
 \braket{\hat{a}^\dagger_0 \hat{\mathcal{C}}_1}_0 
 -\bm{\mathcal{E}}(t) \cdot {\bm{d}^{\eta}}_{\text{0LL}}^{\{1;\mathcal{C}\}} \left[ f \big( E^\eta_{1,\mathcal{C}} \big) - f \big( E^\eta_{\text{0LL}} \big) \right]. 
 \label{eq:EOMa0c1LR}
\end{equation}
Regarding the validity of the approximations, both procedures are consistent with an expansion of the polarization density up to the first order in the electric field and are therefore valid within the linear response theory.

To solve Eq.~\eqref{eq:EOMa0c1LR}, we first express the electric field through its Fourier transform, $\bm{\mathcal{E}}(\omega)$, where $\omega$ is the angular frequency.
Then, considering the adiabatic regime ---meaning that the external fields are switched on very slowly---, we get
\begin{equation}
 \braket{\hat{a}^\dagger_0 \hat{\mathcal{C}}_1}_0  = \int_{\mathbb{R}} \frac{d\omega}{2\pi} \bm{\mathcal{E}}(\omega) \cdot {\bm{d}^{\eta}}_{\text{0LL}}^{\{1;\mathcal{C}\}} \frac{f \big( E^\eta_{1,\mathcal{C}} \big) - f \big( E^\eta_{\text{0LL}} \big)}{E^\eta_{\text{0LL}} - E^\eta_{1,\mathcal{C}} + \hbar \omega} \mathrm{e}^{-\mathrm{i}\omega t}, 
 \label{eq:EOMa0c1sol}
\end{equation}
where we have imposed all averages to be null at $t_0$ ($t_0$ being the initial time in which the perturbation is turned on) and made $t_0 \rightarrow -\infty$, arguing that we have waited long enough for the transient terms to become negligible.
In Eq.~\eqref{eq:EOMa0c1sol}, the substitution $\hbar \omega \rightarrow \hbar \omega + \mathrm{i}\Gamma, \ \Gamma \rightarrow 0^+$ is implied due to the adiabatic limit.
A finite empirical broadening parameter $\Gamma$ is typically considered to account for disorder effects.
As final remark, we stress that this solution is straightforwardly generalizable for all the other pairs of operators.
For example, if we want the expression for $\braket{\hat{\mathcal{C}}^\dagger_n \hat{\mathcal{V}}_{n+1}}_0$, we change from 0LL to $\{n;\mathcal{C}\}$ and from $\{1;\mathcal{C}\}$ to $\{n+1;\mathcal{V}\}$ in the right hand side of Eq.~\eqref{eq:EOMa0c1sol}.

With the previous results, we can write the expectation value of the polarization density operator as
\begin{align}
\braket{\hat{\bm{P}}(t)}_0 &= \sum_\eta \sum_{k_y} \sum_{\substack{\{n;\lambda\} \\ \{n';\lambda'\}}}  \int_{\mathbb{R}} \frac{d\omega}{2\pi} \mathrm{e}^{-\mathrm{i}\omega t}
\frac{f \big( E^\eta_{n',\lambda'} \big) - f \big( E^\eta_{n,\lambda} \big)}{A} \left( {\bm{d}^{\eta}}_{\{n;\lambda\}}^{\{n';\lambda'\}} \right)^*
\frac{\bm{\mathcal{E}}(\omega) \cdot  {\bm{d}^{\eta}}_{\{n;\lambda\}}^{\{n';\lambda'\}}}{E^\eta_{n,\lambda} - E^\eta_{n',\lambda'} + \hbar \omega } \nonumber \\ 
&= \sum_\eta \sum_{\{n;\lambda\}, \lambda'}  \int_{\mathbb{R}} \frac{d\omega}{2\pi} \mathrm{e}^{-\mathrm{i}\omega t} 
\frac{f \big( E^\eta_{n+1,\lambda'} \big) - f \big( E^\eta_{n,\lambda} \big)}{2 \pi l^2_B} \times \nonumber \\
& \quad \times \left[ \left( {\bm{d}^{\eta}}_{\{n;\lambda\}}^{\{n+1;\lambda'\}} \right)^* 
\frac{ \bm{\mathcal{E}}(\omega) \cdot  {\bm{d}^{\eta}}_{\{n;\lambda\}}^{\{n+1;\lambda'\}}}{E^\eta_{n,\lambda} - E^\eta_{n+1,\lambda'} + \hbar \omega } 
+ {\bm{d}^{\eta}}_{\{n;\lambda\}}^{\{n+1;\lambda'\}} 
\frac{\bm{\mathcal{E}}(\omega) \cdot \left( {\bm{d}^{\eta}}_{\{n;\lambda\}}^{\{n+1;\lambda'\}} \right)^* }{E^\eta_{n,\lambda} - E^\eta_{n+1,\lambda'} - \hbar \omega} \right],
\end{align}
where we have performed a trivial summation over $k_y$, which yields the degeneracy of the LLs, $\frac{A}{2 \pi l^2_B}$.
In addition, we clarify that the final expression is obtained employing the optical selection rules and rearranging the summations in a convenient manner.

\section{Demonstration that $\chi_{yx}(\omega)=0$ at half filling}
\label{sec:AppendixB}
We want to prove that 
\begin{equation}
\chi_{yx} (\omega) = \mathrm{i} \sum_\eta \sum_{\{n;\lambda\},\lambda'} \frac{f\left(E_{n+1,\lambda'}^\eta\right) - f\left(E_{n,\lambda}^\eta\right)}{2 \pi l_B^2 \varepsilon_0}
\left| {d_x^{\eta}}_{\{n;\lambda\}}^{\{n+1;\lambda'\}} \right|^2 
\left( \frac{1}{E_{n,\lambda}^\eta - E_{n+1,\lambda'}^\eta + \hbar \omega + \mathrm{i} \Gamma} - 
\frac{1}{E_{n,\lambda}^\eta - E_{n+1,\lambda'}^\eta - \hbar \omega - \mathrm{i} \Gamma} \right)
\end{equation}
vanishes at half filling.

Given that $k_B T \ll 2\Delta$ (even at room temperature), the Pauli exclusion principle implies that only interband transitions are allowed.
As a consequence, we have
\begin{equation}
f\left(E_{n+1,\lambda'}^\eta\right) - f\left(E_{n,\lambda}^\eta\right) = \lambda \delta_{\lambda',-\lambda}.
\end{equation}
Using this result, we can write
\begin{equation}
\chi_{yx} (\omega) = \sum_\eta \chi_{yx}^\eta (\omega), 
\end{equation}
with
\begin{equation}
\chi_{yx}^\eta (\omega) = \mathrm{i} \sum_{\{n;\lambda\}} \frac{\lambda}{2 \pi l_B^2 \varepsilon_0}
\left| {d_x^{\eta}}_{\{n;\lambda\}}^{\{n+1;-\lambda\}} \right|^2 
\left( \frac{1}{E_{n,\lambda}^\eta - E_{n+1,-\lambda}^\eta + \hbar \omega + \mathrm{i} \Gamma} - 
\frac{1}{E_{n,\lambda}^\eta - E_{n+1,-\lambda}^\eta - \hbar \omega - \mathrm{i} \Gamma} \right).
\label{eq:chi_yx_eta}
\end{equation}

In general, $\chi_{yx}^\eta(\omega) $ is not null, meaning that each valley and spin channel yields a finite Hall response.
However, when summing over $\eta$, the contributions cancel out.
In particular, the contribution from $\{\tau;s\}$ cancels out with the one from $\{-\tau;-s\}$, i.e., $\chi_{yx}^{\tau,s} (\omega) = -\chi_{yx}^{-\tau,-s} (\omega)$.
To show this in a rigorous manner, it is helpful to take Eq.~\eqref{eq:chi_yx_eta} and split the sum over LLs in the cases $n=0$, for which $\{n;\lambda\}=\{0;-\tau\}$, and $n \neq 0$, for which the sum runs over $n>0$ and $\lambda=\pm$.
Accordingly, we write
\begin{equation}
\chi_{yx}^\eta (\omega) =  \chi^\eta_{\substack{yx \\ \text{0LL}}} (\omega) + \chi^\eta_{\substack{yx \\ n \neq 0}} (\omega).
\end{equation}
Now, we make use of the identity that relates counterpart transitions, Eq.~\eqref{eq:dipoles_ident}, along with the general relation
\begin{equation}
E_{n,\lambda}^{\tau,s} - E_{n+1,-\lambda}^{\tau,s} = -\left( E_{n,-\lambda}^{-\tau,-s} - E_{n+1,\lambda}^{-\tau,-s} \right),
\end{equation}
to show that
\begin{align}
\chi^{\tau,s}_{\substack{yx \\ \text{0LL}}} (\omega) &= \mathrm{i} \frac{-\tau}{2 \pi l_B^2 \varepsilon_0} \left| {d_x^{\tau,s}}_{\{0;-\tau\}}^{\{1;\tau\}} \right|^2
\left( \frac{1}{E_{0,-\tau}^{\tau,s} - E_{1,\tau}^{\tau,s} + \hbar \omega + \mathrm{i} \Gamma} - 
\frac{1}{E_{0,-\tau}^{\tau,s} - E_{1,\tau}^{\tau,s} - \hbar \omega - \mathrm{i} \Gamma} \right) \nonumber \\
&= \mathrm{i} \frac{\tau}{2 \pi l_B^2 \varepsilon_0} \left| {d_x^{-\tau,-s}}_{\{0;\tau\}}^{\{1;-\tau\}} \right|^2
\left( \frac{1}{E_{0,\tau}^{-\tau,-s} - E_{1,-\tau}^{-\tau,-s} - \hbar \omega - \mathrm{i} \Gamma} - 
\frac{1}{E_{0,\tau}^{-\tau,-s} - E_{1,-\tau}^{-\tau,-s} + \hbar \omega + \mathrm{i} \Gamma} \right) \nonumber \\ 
&= -\chi^{-\tau,-s}_{\substack{yx \\ \text{0LL}}} (\omega), \\
\chi^{\tau,s}_{\substack{yx \\ n \neq 0}} (\omega) &= \mathrm{i} \sum_{n>0,\lambda=\pm} 
\frac{\lambda}{2 \pi l_B^2 \varepsilon_0} \left| {d_x^{\tau,s}}_{\{n;\lambda\}}^{\{n+1;-\lambda\}} \right|^2 
\left( \frac{1}{E_{n,\lambda}^{\tau,s} - E_{n+1,-\lambda}^{\tau,s} + \hbar \omega + \mathrm{i} \Gamma} - 
\frac{1}{E_{n,\lambda}^{\tau,s} - E_{n+1,-\lambda}^{\tau,s} - \hbar \omega - \mathrm{i} \Gamma} \right) \nonumber \\
&= \mathrm{i} \sum_{n>0,\lambda=\pm} 
\frac{-\lambda}{2 \pi l_B^2 \varepsilon_0} \left| {d_x^{\tau,s}}_{\{n;-\lambda\}}^{\{n+1;\lambda\}} \right|^2 
\left( \frac{1}{E_{n,-\lambda}^{\tau,s} - E_{n+1,\lambda}^{\tau,s} + \hbar \omega + \mathrm{i} \Gamma} - 
\frac{1}{E_{n,-\lambda}^{\tau,s} - E_{n+1,\lambda}^{\tau,s} - \hbar \omega - \mathrm{i} \Gamma} \right) \nonumber \\
&= \mathrm{i} \sum_{n>0,\lambda=\pm} 
\frac{\lambda}{2 \pi l_B^2 \varepsilon_0} \left| {d_x^{-\tau,-s}}_{\{n;\lambda\}}^{\{n+1;-\lambda\}} \right|^2 
\left( \frac{1}{E_{n,\lambda}^{-\tau,-s} - E_{n+1,-\lambda}^{-\tau,-s} - \hbar \omega - \mathrm{i} \Gamma} 
- \frac{1}{E_{n,\lambda}^{-\tau,-s} - E_{n+1,-\lambda}^{-\tau,-s} + \hbar \omega + \mathrm{i} \Gamma} \right) \nonumber \\
&= -\chi^{-\tau,-s}_{\substack{yx \\ n \neq 0}} (\omega).
\end{align}

\section{Interacting problem} 
\label{sec:AppendixC}

\subsection*{C.1: Interacting equation of motion} 
\label{subsec:AppendixC1}
The interacting EOM is obtained by adding the result of the commutator with $\hat{H}_{ee} (t)$ in the non-interacting EOM.
As in Appendix~\hyperref[sec:AppendixA]{A}, we present the explicit calculations for only one of the relevant pairs of operators, $\hat{c}^\dagger_{\{ \text{0LL}; k_y \},\eta}(t) \hat{c}_{\{1;\mathcal{C}; k_y \},\eta}(t) \equiv \hat{a}^\dagger_0 \hat{\mathcal{C}}_1$.
The derivation for the other pairs follows analogously.

After some straightforward algebra, we get
\begin{equation}
\left[\hat{H}_{ee}(t), \hat{a}^\dagger_0 \hat{\mathcal{C}}_1 \right] = \&_1 + \&_2 + \&_3 + \&_4, 
\label{eq:commutatorHee}
\end{equation}
where
\begin{equation}
\&_1 = \frac{1}{2} \sum_{s''} \sum_{\substack{\alpha_1, \alpha_2 \\ \alpha_3}}
U^{\tau,s,s''}_{\substack{\alpha_1, \alpha_2 \\ \alpha_3, \{ \text{0LL};k_y \}}}
\hat{c}^\dagger_{\alpha_1,\tau,s} (t) \hat{c}^\dagger_{\alpha_2,\tau,s''} (t)\hat{c}_{\alpha_3,\tau,s''} (t) \hat{\mathcal{C}}_1,
\end{equation}
\begin{equation}
\&_2 = -\frac{1}{2} \sum_{s'} \sum_{\substack{\alpha_1, \alpha_2 \\ \alpha_4}}
U^{\tau,s',s}_{\substack{\alpha_1, \alpha_2 \\ \{ \text{0LL};k_y \}, \alpha_4}}
\hat{c}^\dagger_{\alpha_1,\tau,s'} (t) \hat{c}^\dagger_{\alpha_2,\tau,s} (t)\hat{c}_{\alpha_4,\tau,s'} (t) \hat{\mathcal{C}}_1, 
\end{equation}
\begin{equation}
\&_3 = \frac{1}{2} \sum_{s'} \sum_{\substack{\alpha_1 \\ \alpha_3, \alpha_4}}
U^{\tau,s',s}_{\substack{\alpha_1, \{ 1;\mathcal{C};k_y \} \\ \alpha_3, \alpha_4}}
\hat{a}^\dagger_0 \hat{c}^\dagger_{\alpha_1,\tau,s'} (t) \hat{c}_{\alpha_3,\tau,s} (t) \hat{c}_{\alpha_4,\tau,s'} (t),
\end{equation}
\begin{equation}
\&_4 = -\frac{1}{2} \sum_{s''} \sum_{\substack{\alpha_2 \\ \alpha_3, \alpha_4}}
U^{\tau,s,s''}_{\substack{\{ 1;\mathcal{C};k_y \}, \alpha_2 \\ \alpha_3, \alpha_4}}
\hat{a}^\dagger_0 \hat{c}^\dagger_{\alpha_2,\tau,s''} (t) \hat{c}_{\alpha_3,\tau,s''} (t) \hat{c}_{\alpha_4,\tau,s} (t).
\end{equation}

\subsection*{C.2: Exchange self-energy terms} 
\label{subsec:AppendixC2}
The interacting EOM contains four new types of terms, as seen in Eq.~\eqref{eq:commutatorHee}.
We first deal with $\&_1$.

Likewise the non-interacting case, it is implicit that, within the linear response limit, we take the average of the new terms with respect to the unperturbed Hamiltonian.
The average of $\&_1$ implies the average of the product of four fermionic operators which, within the random phase approximation, yields
\begin{equation}
\braket{\hat{c}^\dagger_{\alpha_1,\tau,s} (t) \hat{c}^\dagger_{\alpha_2,\tau,s''} (t)
\hat{c}_{\alpha_3,\tau,s''} (t) \hat{\mathcal{C}}_1}_0 = 
\braket{\hat{c}^\dagger_{\alpha_1,\tau,s} (t) \hat{\mathcal{C}}_1}_0
\braket{\hat{c}^\dagger_{\alpha_2,\tau,s''} (t) \hat{c}_{\alpha_3,\tau,s''} (t)}_0 
-\braket{\hat{c}^\dagger_{\alpha_1,\tau,s} (t) \hat{c}_{\alpha_3,\tau,s''} (t)}_0
\braket{\hat{c}^\dagger_{\alpha_2,\tau,s''} (t) \hat{\mathcal{C}}_1}_0.
\end{equation}
Among these terms, the ones that lead to a band renormalization ---the so-called SE terms--- are
\begin{equation}
\braket{\hat{c}^\dagger_{\alpha_1,\tau,s} (t) \hat{c}^\dagger_{\alpha_2,\tau,s''} (t)
\hat{c}_{\alpha_3,\tau,s''} (t) \hat{\mathcal{C}}_1}^\text{SE}_0 = 
\delta_{\alpha_1,\{\text{0LL},k_y\}} \delta_{\alpha_2,\alpha_3} 
\braket{\hat{a}^\dagger_0 \hat{\mathcal{C}}_1}_0 f\left( E^{\tau,s''}_{\alpha_2} \right) 
- \delta_{s,s''} \delta_{\alpha_1,\alpha_3} \delta_{\alpha_2,\{\text{0LL},k_y\}}
\braket{\hat{a}^\dagger_0 \hat{\mathcal{C}}_1}_0 f\left( E^{\tau,s}_{\alpha_1} \right).
\end{equation}
This leads to
\begin{equation}
\braket{\&_1}^\text{SE}_0 =  \braket{\&_1}^\text{Hartree}_0 + \braket{\&_1}^\text{Fock}_0,
\end{equation}
where
\begin{equation}
\braket{\&_1}^\text{Hartree}_0 = \frac{1}{2} \braket{\hat{a}^\dagger_0 \hat{\mathcal{C}}_1}_0 
\sum_{s''} \sum_{\alpha_2} U^{\tau,s,s''}_{\substack{\{ \text{0LL};k_y \}, \alpha_2 \\ \alpha_2, \{ \text{0LL};k_y \}}} f\left( E^{\tau,s''}_{\alpha_2} \right) 
\end{equation}
is the Hartree term and 
\begin{equation}
\braket{\&_1}^\text{Fock}_0 = - \frac{1}{2} \braket{\hat{a}^\dagger_0 \hat{\mathcal{C}}_1}_0 
\sum_{\alpha_1} U^{\tau,s,s}_{\substack{\alpha_1, \{ \text{0LL};k_y \} \\ \alpha_1, \{ \text{0LL};k_y \}}} f\left( E^{\tau,s}_{\alpha_1} \right) 
\end{equation}
is the Fock or exchange SE term.

By analogy with the Hartree-Fock approximation to the problem of the homogeneous electron gas~\cite{Mahan2013}, we argue that the Hartree term, which mixes spins, is canceled by the electron-ion background within the Jellium model.
To support this claim, we have verified that the limit $B_0 = 0$ in $\braket{\&_1}^\text{Hartree}_0$ implies a null transferred momentum, i.e., $\delta_{\bm{q},\bm{0}}$.
As a result, we keep only the Fock term, which couples the same spin flavors.

Repeating the same calculations, and making use of the identity
\begin{equation}
U^{\tau,s,s'}_{\substack{\alpha_1, \alpha_2 \\ \alpha_3, \alpha_4}} = U^{\tau,s',s}_{\substack{\alpha_2, \alpha_1 \\ \alpha_4, \alpha_3}},
\end{equation}
it is immediate to show that $\braket{\&_2}^\text{Fock}_0 = \braket{\&_1}^\text{Fock}_0$, which leads to
\begin{equation}
\braket{\&_1 + \&_2}^\text{Fock}_0 = - \braket{\hat{a}^\dagger_0 \hat{\mathcal{C}}_1}_0 
\sum_{\alpha_1} U^{\tau,s,s}_{\substack{\alpha_1, \{ \text{0LL};k_y \} \\ \alpha_1, \{ \text{0LL};k_y \}}} f\left( E^{\tau,s}_{\alpha_1} \right).
\end{equation}
Similarly, we obtain 
\begin{equation}
\braket{\&_3 + \&_4}^\text{Fock}_0 = \braket{\hat{a}^\dagger_0 \hat{\mathcal{C}}_1}_0 
\sum_{\alpha_1} U^{\tau,s,s}_{\substack{\alpha_1, \{ 1;\mathcal{C};k_y \} \\ \alpha_1, \{ 1;\mathcal{C};k_y \}}} f\left( E^{\tau,s}_{\alpha_1} \right).
\end{equation}

We now observe that the interacting EOM is equivalent to the non interacting one, Eq.~\eqref{eq:EOMa0c1LR}, with a renormalized energy difference, given by
\begin{equation}
\left( E^\eta_{\text{0LL}} - E^\eta_{1,\mathcal{C}} \right)_\text{renorm} = E^\eta_{\text{0LL}} - E^\eta_{1,\mathcal{C}} 
+ \sum_{\alpha_1} f\left( E^{\tau,s}_{\alpha_1} \right) 
\left[ U^{\tau,s,s}_{\substack{\alpha_1, \{ 1;\mathcal{C};k_y \} \\ \alpha_1, \{ 1;\mathcal{C};k_y \}}} 
-U^{\tau,s,s}_{\substack{\alpha_1, \{ \text{0LL};k_y \} \\ \alpha_1, \{ \text{0LL};k_y \}}} \right].
\end{equation}

Generalizing these results for the other pairs of operators, we conclude that the energy bands are renormalized as 
\begin{equation}
\left( E^\eta_\alpha \right)_{\text{renorm}} = E^\eta_\alpha + \Sigma^\eta_\alpha,
\end{equation}
where
\begin{equation}
\Sigma^\eta_\alpha = - \sum_{\alpha'} f\left( E^\eta_{\alpha'} \right) U^{\tau,s,s}_{\substack{\alpha', \alpha \\ \alpha', \alpha}} 
\label{eq:SEappendix}
\end{equation}
are the exchange SE corrections.

\subsection*{C.3: Coulomb integrals} 
\label{subsec:AppendixC3}
The (general) expression for the exchange SE corrections, Eq.~\eqref{eq:SEappendix}, hides multiple integrals that can be solved analytically. 
Here, we provide some of the technical steps that lead to the simplification of this expression.

We turn our attention to the following integral,
\begin{equation}
I_0 \equiv \int_{-\infty}^{+\infty} dx \ \mathrm{e}^{\mathrm{i}q_x x} 
\frac{ \mathrm{e}^{-\left( \frac{x}{l_B} + l_B k'_y \right)^2/2} \ \mathrm{e}^{-\left( \frac{x}{l_B} + l_B k_y \right)^2/2} }{\sqrt{\pi} l_B} 
\tilde{H}_{n'} \left( \frac{x}{l_B} + l_B k'_y \right) \tilde{H}_n \left( \frac{x}{l_B} + l_B k_y \right).
\end{equation}
This integral is relevant as its solution includes the non-trivial steps required to calculate the structure factors that lie inside the Coulomb integrals (see Eqs.~\eqref{eq:Coulomb_integrals}-\eqref{eq:structure_factors}).

With the change of variables $u=\frac{x}{l_B} + l_B k'_y$, we obtain
\begin{equation}
I_0 =  \mathrm{e}^{-l^2_B \left( q^2_y/2 + \mathrm{i} q_x k'_y \right)} I_1,
\end{equation}
with
\begin{equation}
I_1 = \int_{-\infty}^{+\infty} \frac{du}{\sqrt{\pi}} \ \mathrm{e}^{-u^2 + l_B \left(q_y + \mathrm{i}q_x \right) u}  
\tilde{H}_{n'} (u) \tilde{H}_n \left( u - l_B q_y \right),
\end{equation}
where $q_y = k'_y - k_y$ is an implicit relation that comes from the trivial integration over $dy$ in the structure factors.

At this point, we resort to a table of integrals, Ref.~\onlinecite{Gradshteyn2014}, and invoke Eq. 7.377 which, with little manipulation, can be written as
\begin{align}
&\int_{-\infty}^{+\infty} \frac{du}{\sqrt{\pi}} \ \mathrm{e}^{-u^2 + l_B \left(q_y + \mathrm{i}q_x \right) u}
\tilde{H}_n \left(u-\frac{l_B (q_y + \mathrm{i} q_x)}{2} + p_0\right) \tilde{H}_{n'} \left( u-\frac{l_B (q_y + \mathrm{i} q_x)}{2} + q_0 \right) \nonumber \\
& \quad = \mathrm{e}^{l^2_B \left(q_y + \mathrm{i}q_x \right)^2/4}  \sqrt{2^{n'-n}} \sqrt{\frac{n!}{n'!}} q_0^{n'-n} L_n^{n'-n} (-2 p_0 q_0 ), \quad [n \leq n'].
\end{align}
Given this relation, it is straightforward to show that
\begin{equation}
I_0 = \mathrm{e}^{-l^2_B \left[\frac{q_x^2 + q_y^2}{4} + \mathrm{i} q_x \left( k_y'-\frac{q_y}{2} \right) \right]} 
\sqrt{2^{|n'-n|}} \sqrt{\frac{\text{min}(n,n')!}{\text{max}(n,n')!}} 
\left[l_B \frac{\text{sign}(n'-n) q_y + \mathrm{i} q_x}{2} \right]^{|n'-n|} L_{\text{min}(n,n')}^{|n'-n|} \left( l_B^2 \frac{q_x^2 + q_y^2}{2} \right).
\end{equation}

The remaining steps required to simplify Eq.~\eqref{eq:SEappendix} into Eqs.~\eqref{eq:SEsimp}-\eqref{eq:SEintegrals} follow straightforwardly.

\end{widetext}

\bibliographystyle{apsrev4-1}

\end{document}